# DISCRIMINATION IN THE AGE OF ALGORITHMS

February 5, 2019

Jon Kleinberg[+]
Jens Ludwig[*]
Sendhil Mullainathan[**]
Cass R. Sunstein[***]

## ABSTRACT

*The law forbids discrimination. But the ambiguity of human decision-making often makes it extraordinarily hard for the legal system to know whether anyone has actually discriminated. To understand how algorithms affect discrimination, we must therefore also understand how they affect the problem of detecting discrimination. By one measure, algorithms are fundamentally opaque, not just cognitively but even mathematically. Yet for the task of proving discrimination, processes involving algorithms can provide crucial forms of transparency that are otherwise unavailable. These benefits do not happen automatically. But with appropriate requirements in place, the use of algorithms will make it possible to more easily examine and interrogate the entire decision process, thereby making it far easier to know whether discrimination has occurred. By forcing a new level of specificity, the use of algorithms also highlights, and makes transparent, central tradeoffs among competing values. Algorithms are not only a threat to be regulated; with the right safeguards in place, they have the potential to be a positive force for equity.*

[+] Tisch University Professor, Cornell University.
[*] Edwin A. and Betty L. Bergman Distinguished Service Professor, University of Chicago.
[**] Roman Family University Professor of Computation and Behavioral Science, University of Chicago.
[***] Robert Walmsley University Professor, Harvard University.

Thanks to Michael Ridgway for his assistance with data analysis; to Justin McCrary for helpful discussions; to Solon Barocas, James Grenier, Saul Levmore, Karen Levy, Eric Posner, Manish Raghavan, and David Robinson for valuable comments; to the MacArthur, Simons, and Russell Sage Foundations for financial support for this work on algorithmic fairness; to the Program on Behavioral Economics and Public Policy at Harvard Law School; and to Tom and Susan Dunn, Ira Handler and the Pritzker Foundation for support of the University of Chicago Urban Labs more generally. Thanks to Andrew Heinrich and Cody Westphal for superb research assistance. All opinions and any errors are obviously ours alone.

# I. INTRODUCTION

The law forbids discrimination, but it can be exceedingly difficult to find out whether human beings have discriminated.[1] Accused of violating the law, people might well dissemble. Some of the time, they themselves might not even be aware that they have discriminated. Human decisions are frequently opaque to outsiders, and they may not be much more transparent to insiders. A defining preoccupation of discrimination law, to which we shall devote considerable attention, is how to handle the resulting problems of proof.[2] Those problems create serious epistemic challenges, and they produce predictable disagreements along ideological lines.

Our central claim here is that when algorithms are involved, proving discrimination will be easier – or at least it should be, and can be made to be. The law forbids discrimination by algorithm, and that prohibition can be implemented by regulating the process through which algorithms are designed. This implementation could codify the most common approach to building machine learning classification algorithms in practice, and add detailed recordkeeping requirements. Such an approach would provide valuable transparency about the decisions and choices made in building algorithms – and also about the tradeoffs among relevant values.

We are keenly aware that these propositions are jarring, and that they will require considerable elaboration. They ought to jar because in a crucial sense algorithms are not decipherable – one cannot determine what an algorithm will do by reading the underlying code. This is more than a cognitive limitation; it is a mathematical impossibility. To know what an algorithm will do, one must run it.[3] The task at hand, though, is to take an observed gap, such as differences in hiring rates by gender, and to decide whether the gap should be attributed to discrimination as the law defines it. Such attributions need not require that we read the code. Instead, they can be accomplished by examining the data given to the algorithm and probing its outputs, a process that (we will argue) is eminently feasible. The opacity of the algorithm does not prevent us from scrutinizing its construction or experimenting with its behavior – two activities that are impossible with humans.[4]

Crucially, policy changes are needed to realize these benefits, such as the requirement that all the components of an algorithm (including the training data) must be stored and made available for examination and experimentation. It is important to see that without the appropriate safeguards, the prospects for detecting discrimination in a world of unregulated algorithm design could become even more serious than they currently are.

---

[1] For an instructive account in the constitutional context, *see* David A. Strauss, *Discriminatory Intent and the Taming of Brown*, 56 U. CHI. L. REV.935 (1989). On the general problem, *see* Audrey Lee, *Unconscious Bias Theory in Employment Discrimination Litigation*, 40 HARV. C.R.-C.L. L. REV. 483 (2005).

[2] *See, e.g.*, McDonnell Douglas Corp. v. Green, 411 U.S. 792 (1973); Watson v. Fort Worth Bank & Trust, 487 U.S. 977 (1988).

[3] *See* Michael Sipser, *Introduction to the Theory of Computation* (2012). For a discussion of some of these issues in a legal setting, see also Deven R. Desai and Joshua A. Kroll, *Trust But Verify: A Guide to Algorithms and the Law*, 31 HARV. J.L. & TECH. 1 (2017).

[4] We employ this dichotomy – processes where decisions are made by algorithms and ones where decisions are made by humans – to make the issues clear. In practice, there can be a hybrid, with humans overriding algorithmic judgments with their own. These hybrid processes will clearly have the two elements of each process we describe here, with the additional observation that, with proper regulation, we can see the exact instances where humans overrode the algorithm.

Our starting point is the current American legal system that has developed a complex framework for detecting and regulating discriminatory decisions.[5] It is increasingly clear that this framework must be adapted for regulating the growing number of questions – involving hiring, credit, admissions, criminal justice – where algorithms are now involved in how public and private institutions decide.[6] Algorithms provide new avenues for people to incorporate past discrimination, or to express their biases and thereby to exacerbate discrimination. Getting the proper regulatory system in place does not simply limit the possibility of discrimination from algorithms; it has the potential to turn algorithms into a powerful counterweight to human discrimination and a positive force for social good of multiple kinds.

We aim here to explore the application of discrimination law to a particularly important category of decisions: *screening decisions*, where a person (or set of people) is chosen from a pool of candidates to accomplish a particular goal, as when college students are admitted on the basis of academic potential or defendants are jailed on the basis of flight risk. [7] Algorithms can be used to produce predictions of the candidate's outcomes, such as future performance after acceptance of a job offer or admission to an academic program. We focus on one kind of machine learning algorithm often applied to such problems, which uses training data to produce a function that takes inputs (such as the characteristics of an applicant) and produces relevant predictions.[8] The terminology here can be confusing since there are actually two algorithms: one algorithm (the '*screener*') that for every potential applicant produces an evaluative score (such as an estimate of future performance); and another algorithm (the '*trainer*') that uses data to produce the screener that best optimizes some objective function.[9] The distinction is important and often overlooked; we shall emphasize it here.

The existing legal framework for these types of screening decisions is necessarily shaped by practical considerations involving the typical difficulty of uncovering human motivations. Simply knowing that there is a disparity in hiring outcomes is not enough to determine whether anyone has discriminated.[10] Perhaps there are genuine differences in average performance across groups.[11] A central aspect of the legal challenge is to determine how and why the hiring decisions were made and whether protected personal characteristics, such as race and gender, played a role.[12] Deciding whether there has been discrimination is difficult for one obvious and one less obvious reason. The obvious reason is generic to any legal system: people dissemble, obfuscate, and lie. The less obvious reason is that people may not even know themselves.

A large body of research from behavioral science, described below, tells us that people themselves may not know why and how they are choosing – even (or perhaps especially) when they think that they do.[13]

---

[5] For a clear, recent treatment, *see* Solon Barocas & Andrew D. Selbst, *Big Data's Disparate Impact*, 104 CAL. L. REV. 671 (2016).

[6] This point has been made by a large and growing literature in computer science. While the literature is vast, some canonical papers include Cynthia Dwork et al., *Fairness Through Awareness*, PROCEEDINGS OF THE 3RD INNOVATIONS IN THEORETICAL COMPUTER SCIENCE CONFERENCE 214, 214-26 (2012); Solon Barocas & Andrew Selbst, *Big Data's Disparate Impact*, 104 CALIF. L. REV. 671 (2016) and the curated set of studies assembled at *Scholarship*, FAIRNESS, ACCOUNTABILITY, AND TRANSPARENCY IN MACHINE LEARNING, https://www.fatml.org/resources/relevant-scholarship (last visited Jan. 14, 2019).

[7] For one example, *see* Jon Kleinberg et al., *Human Decisions and Machine Predictions*, 133 Q.J. ECON. 237 (2018).

[8] *See id*.

[9] There are clearly other kinds of algorithms and decisions, and they will require an independent analysis.

[10] *See* McDonnell Douglas Corp. v. Green, 411 U.S. 792 (1973).

[11] *See* Lee, *supra* note; Tristin K. Green, *Making Sense of the* McDonnell Douglas *Framework: Circumstantial Evidence and Proof of Disparate Treatment under Title VII*, 87 CALIF. L. REV. 983 (1999).

[12] *See, e.g.*, Corning Glass Works v. Brennan, 417 US 188 (1974).

[13] *See* Richard E. Nisbett & Timothy DeCamp Wilson, *Telling More Than We Can Know: Verbal Reports on Mental Processes*, 84 PSYCHOL. REV. 231 (1977).

Many choices happen automatically; the influences of choice can be subconscious; and the rationales we produce are constructed after the fact and on the fly.[14] This means that witnesses and defendants may have difficulty accurately answering the core questions at the heart of most discrimination cases: What screening rule was used? And why? Even the most well-intentioned people may possess unconscious or implicit biases against certain groups.[15]

In contrast, a well-regulated process involving algorithms stands out for its *transparency* and *specificity:* it is not obscured by the same haze of ambiguity that obfuscates human decision-making. Access to the algorithm allows us to ask questions that we cannot meaningfully ask of human beings. For any candidate, we can ask: "How would the screening rule's decision have been different if a particular feature (or features) of the applicant were changed?" We can ask exactly which data were made available for training the algorithm (and which were not), as well as the precise objective function that was maximized during the training. We will show that, as a result, we can attribute any observed disparity in outcomes to the different components of the algorithm design or conclude that the disparity is due to structural disadvantages outside of the decision process. In a nutshell: For the legal system, discovering "on what basis are they choosing?" and "why did they pick those factors?" becomes much more feasible.

It would be naïve – even dangerous – to conflate "algorithmic" with "objective," or to think that the use of algorithms will necessarily eliminate discrimination against protected groups.[16] The reliance on data does not provide algorithms a presumption of truth; the data they are fed can be biased, perhaps because they are rooted in past discrimination (as, for example, where past arrest records are used to predict the likelihood of future crime).[17] It would also be naive to imagine that the specificity of algorithmic code does not leave room for ambiguity elsewhere. Algorithms do not build themselves. The Achilles' heel of all algorithms is the humans who build them and the choices they make about outcomes, candidate predictors for the algorithm to consider, and the training sample. A critical element of regulating algorithms is regulating humans.[18] Algorithms change the landscape – they do not eliminate the problem.

## A. Implications of Our Framework

Five points are central to our analysis of discrimination law in an age of algorithms. First, the challenge of regulating discrimination is fundamentally one of *attribution*. When a screening process produces a disparity for a particular group, to what do we attribute that gap? It could come from the screening rule used. Perhaps the screening rule explicitly takes account of gender. Perhaps the chosen objective – the outcome the screening rule aims to optimize – disadvantages women (or some other protected group). Perhaps the disparity comes from the set of inputs made available to the screener. Perhaps the screening rule fails to optimize for a given outcome using the inputs. The answers to these questions may or may not be relevant, depending on what the pertinent law considers to be "discrimination."

Importantly, the disparity may not result from any of these problems with the screening rule itself. It could also be the consequence of average differences in the outcome distributions across groups. For example, if some groups truly have worse access to K-12 schooling opportunities – perhaps their schools have lower levels of resources – then their college application packets may be less strong on average. Decomposing the source of disparities in screening decisions can be enormously difficult, but it is critical for determining when legal remedies should be applied, and which ones.

[14] *See* Timothy Wilson, *Strangers to Ourselves* (2004).
[15] *See* Mahzarin Banjali & Anthony Greenwald, *Blind Spots* (2013).
[16] *See* Barocas & Selbst, *supra* note.
[17] *See id.*; Sandra G. Mayson, *Bias In, Bias Out*, 128 YALE L.J. (forthcoming 2019), available at https://papers.ssrn.com/sol3/papers.cfm?abstract_id=3257004; Sharad Goel et al., *The Accuracy, Equity, and Jurisprudence of Criminal Risk Assessment* (2019), available at https://papers.ssrn.com/sol3/papers.cfm?abstract_id=3306723.
[18] As another example, in some cases, humans may choose to override the algorithm and this re-introduces human ambiguity. It is worth noting here that we can see when these overrides arise.

Second, this decomposition becomes easier once an algorithm is in the decision loop. Now the decisions we examine are far more specific than "why was this particular candidate chosen?" For example, a key input into the training algorithm is a choice of objective – given the data, the trainer must produce a screening rule that identifies people predicted to do well on some outcome (for example, salespeople expected to have the highest revenues generated). Algorithms are *exceedingly sensitive* to these choices.[19] In searching for discrimination, the legal system may or may not make it important to ask, "Was the training algorithm given an appropriate outcome to predict?" The ability to ask this question is a luxury: Instead of trying to infer why a salesperson was hired, the algorithm's objective function provides us with such information.

The luxury of this knowledge unlocks the power of scrutiny: was this a reasonable choice of outcome?[20] The same point holds for the other key choices for the trainer. Of course, the ability to obtain the relevant knowledge requires (and we argue for) a high degree of transparency. At a minimum, these records and data should be stored for purposes of discovery. Algorithms do not only provide the means to scrutinize the choices we make in building them, they *demand* that scrutiny: it is with respect to these choices that human bias can creep into the algorithm.

Third, such scrutiny pays a dividend: if we regulate the human choices well, we might be willing to be more permissive towards how the algorithm uses information about personal attributes in certain ways. When we ensure human choices are made appropriately, some of the concerns that animate the existing legal framework for human discrimination are rendered moot for the algorithm. Suppose, for example, that college applications require recommendations from high school teachers. Any racial bias by teachers could lead to differences in average letter quality across race groups. Interestingly, in some cases the best way to mitigate the discriminatory effects of biased data is *to authorize the algorithm to have access to information about race*.[21] To see why, note that *only* an algorithm that sees race can detect that someone from a given group has better college grades than their letters would suggest, and then adjust predicted performance to address this disparity. Yet much of the time, considerations of factors like race is what antidiscrimination law seeks to prevent (though in this setting, the legal result is not entirely clear).[22]

Fourth, algorithms will force us to make more explicit judgments about underlying principles. If our goals are in tension – as, for example, if admitting more minority students into an elite college would reduce first-year GPAs because of disparities in K-12 school quality or other structural disadvantages – the algorithm precisely quantifies this tradeoff. And we must now articulate a choice. What tradeoff do we find acceptable? We will now be in a position to grapple with such questions *quantitatively*.[23]

---

[19] *See* Kleinberg et al., *supra* note.

[20] We are bracketing, for the moment, the question whether that is legally relevant.

[21] Jon Kleinberg, Jens Ludwig et al., *Algorithmic Fairness*, 108 AM. ECON. REV. PAPERS & PROCEEDINGS 22 (2018). See also Talia Gillis and Jann Spiess, *Big Data and Discrimination*, U. Chi. L. Rev. (forthcoming 2019), available at https://papers.ssrn.com/sol3/papers.cfm?abstract_id=3204674.

[22] *See* Loving v. Virginia, 388 U.S. 1 (1967); Miller v. Johnson, 515 U.S. 800, 811 (1995). One possible response to this example is to argue that if the data contain bias, we simply should not use them at all. But in many applications, it is difficult to imagine any alternative to data-driven decision-making. In the case of mortgage lending, for instances, absent information about the current income or assets of a borrower, or prior repayment history, on what basis should a lender decide risk? A middle-ground approach might be only to use those data that are not so biased, but as argued above, the algorithm has a much greater chance of being able to tell which data elements contain a differential signal for one group relative to another than would any human being.

Our fifth and final point is that if appropriate regulation can protect against malfeasance in their deployment, then algorithms can become a potentially powerful force for good: they can dramatically reduce discrimination of multiple kinds. A variety of research shows that unstructured decision-making is exactly the sort of environment in which implicit biases can have their biggest impact.[24] Of course it is true that in building an algorithm, human beings can introduce biases in their choice of objectives and data; importantly, they might use data that are themselves a product of discrimination. But conditional on getting objectives and data right, the algorithm at least removes the human bias of an unstructured decision process. The algorithm, unlike the human being, has no intrinsic preference for discrimination, and no ulterior motives.

And this might not even be the source of the most important gains for disadvantaged groups. In many contexts, efficiency improvements alone have large disparate benefits for members of such groups. For example, Kleinberg et al. examine pre-trial release decisions in New York, and find that algorithms better distinguish low-risk from high-risk defendants.[25] By prioritizing the highest-risk people to detain, it becomes feasible in principle to jail 42% fewer people with no increase in crime.[26] The biggest benefits would accrue to the two groups that currently account for nine of every ten jail inmates: African-Americans and Hispanics.

We develop these points at length, beginning with an exploration of discrimination law, the relevance of principles from behavioral science, and the tensions introduced into this framework by the use of algorithms.[27] Our central claim, stated in simple form, is that we do not need safeguards against algorithms; we need safeguards against the biases of the people who built them. We must scrutinize and regulate those choices, to ensure that algorithms are not being built in a way that discriminates (recognizing the complexity and contested character of that term). If we do that, then algorithms go beyond merely being a threat to be regulated; they can also be a positive force for social justice.

## II. The Law of Discrimination: A Primer

Discrimination law has long been focused on two different problems. The first is *disparate treatment*; the second is *disparate impact*. The Equal Protection Clause of the Constitution,[28] and all civil rights laws, forbid disparate treatment.[29] The Equal Protection Clause of the Constitution does not concern itself with disparate impact,[30] but some civil rights statutes do.[31]

---

[23] See *infra* for details.
[24] For evidence, see Crystal S. Yang, *Free at Last? Judicial Discretion and Racial Disparities in Federal Sentencing*, 44 J. LEGAL STUD. 75 (2015); Alma Cohen & Crystal S. Yang, *Judicial Politics and Sentencing Decisions*, AM. ECON. J.: ECON. POL'Y (forthcoming 2018).
[25] *See* Jon Kleinberg et al., *Human Decisions and Machine Predictions*, 133 Q.J. ECON 237 (2018).
[26] *Id.*
[27] We are building on an emerging line of research connected to developments in computer science, including valuable recent work by Barocas and Selbst that seeks to situate algorithms within the framework of discrimination law. *See* Barocas & Selbst, *supra* note.
[28] *See* Vasquez v. Hillery, 474 US 254 (1986).

***Disparate treatment***. The prohibition on disparate treatment reflects a commitment to a kind of neutrality.[32] For example, public officials are not permitted to favor men over women or white people over black people. Civil rights statutes forbid disparate treatment along a variety of specified grounds, such as race, sex, national origin, religion, and age.[33]

In extreme cases, the existence of disparate treatment is obvious, because a facially discriminatory practice or rule can be shown to be in place ("no women may apply").[34] In other cases, no such practice or rule can be identified, and for that reason, violations are more difficult to police.[35] A plaintiff might claim that a facially neutral practice or requirement (such as a written test for employment) was actually adopted in order to favor one group (whites) or to disfavor another (Hispanics).[36] To police discrimination, the legal system is required to use what tools it has to discern the motivation of decision-makers.[37] To paraphrase the Supreme Court, the key question under the Equal Protection Clause is simple: *Was the requirement or practice chosen because of, rather than in spite of, its adverse effects on relevant group members?*[38] That question might be exceedingly challenging to answer, but the law makes it necessary to try.[39]

It is important to see that the disparate treatment idea applies whether discrimination is taste-based or statistical.[40] An employer might discriminate (1) because he himself prefers working with men to working with women; (2) because the firm's coworkers prefer working with men to working with women; or (3) because customers prefer men in the relevant positions. In all of these cases, disparate treatment is strictly forbidden.[41] Or suppose that an employer is relying on a statistical demonstration that (for example) women leave the workforce more frequently than men do, or that women over 50 are more likely than men over 50 to leave within ten years. Even if the statistical demonstration is convincing, facial discrimination against women, or some kind of boost for men or penalty for women, is strictly prohibited.[42]

---

[29] For helpful discussion, *see* Cary Franklin, *Inventing the "Traditional Concept" of Sex Discrimination*, 125 HARV. L. REV. 1309 (2012); Elizabeth Bartholet, *Application of Title VII to Jobs in High Places*, 95 HARV. L. REV. 945 (1982); Miguel Mendez, *Presumptions of Discriminatory Motive in Title VII Disparate Treatment Cases*, 32 STAN. L. REV. 1129 (1980).
[30] *See* Washington v. Davis, 426 US 229 (1976); McCleskey v. Kemp, 481 US 279 (1987).
[31] *See, e.g.,* Griggs v. Duke Power Co., 401 U.S. 424 (1971); Meacham v. Knolls Atomic Power Lab., 554 U.S. 84 (2008). In the context of age discrimination, *see* Smith v. City of Jackson, 544 U.S. 229 (2005). For discussion, *see* George Rutherglen, *Disparate Impact, Discrimination, and the Essentially Contested Concept of Equality*, 74 FORDHAM L. REV. 2313 (2006); Richard A. Primus, *Equal Protection and Disparate Impact: Round Three*, 117 HARV. L. REV. 493 (2004); Samuel R. Bagenstos, *The Structural Turn and the Limits of Antidiscrimination Law*, 94 CALIF. L. REV. 1, 4 (2006); Michael Selmi, *Was the Disparate Impact Theory a Mistake?*, 53 UCLA L. REV.. 701, 732–45 (2006). For objections, see Michael Evan Gold, Griggs' *Folly: An Essay on the Theory, Problems, and Origin of the Adverse Impact Definition of Employment Discrimination and a Recommendation for Reform*, 7 INDUS. REL. L.J. 429, 491–500 (1985).
[32] *See* Paul Brest, *In Defense of the Antidiscrimination Principle*, 90 HARV L. REV. 1 (1976).
[33] *See, e.g.,* 42 U.S.C. § 2000e-2 ("It shall be an unlawful employment practice for an employer -
(1) to fail or refuse to hire or to discharge any individual, or otherwise to discriminate against any individual with respect to his compensation, terms, conditions, or privileges of employment, because of such individual's race, color, religion, sex, or national origin.").
[34] Reed v. Reed, 404 U.S. 71 (1971).
[35] *See* Bartholet, *supra* note; Christine Jolls & Cass R. Sunstein, *The Law of Implicit Bias*, 94 CALIF. L. REV. 969 (2006).
[36] Washington v. Davis, 426 U.S. 229 (1976).
[37] A defining framework can be found in *McDonnell Douglas Corp. v. Green*, 411 U.S. 792 (1973). *See* Tristin K. Green, *Making Sense of the* McDonnell Douglas *Framework: Circumstantial Evidence and Proof of Disparate Treatment under Title VII*, 87 CAL. L. REV. 983 (1999).
[38] Personnel Adm'r of Massachusetts v. Feeney, 442 U.S. 256 (1979).
[39] We are bracketing some of the differences between the Equal Protection Clause and the civil rights statutes. On the latter, see Green, *supra* note.
[40] The classic discussion of taste-based discrimination is Gary Becker, *The Economics of Discrimination* (1971). On statistical discrimination, *see* Edmund Phelps, *The Statistical Theory of Racism and Sexism*, 62 AM. ECON. REV. 659 (1972). On the difference, *see* Jonathan Guryan & Kerwin Kofi Charles, *Taste-Based or Statistical Discrimination: The Economics of Discrimination Returns to Its Roots*, 123 ECON. J. F417 (2013); Cass R. Sunstein, *Why Markets Won't Stop Discrimination*, 8 SOC. PHIL. & POL'Y 22 (1991).

All this is relatively straightforward as a matter of law, but as a matter of principle, it raises some serious puzzles, to which we will return.

***Disparate impact***. The prohibition on disparate impact means, in brief, that if some requirement or practice has a disproportionate adverse effect on members of protected groups (such as women and African-Americans), the defendant must show that the requirement or practice is adequately justified.[43] Suppose, for example, that an employer requires members of its sales force to take some kind of written examination, or that the head of a police department institutes a rule requiring new employees to be able to run at a specified speed. If these practices have disproportionate adverse effects on members of protected groups, they will be invalidated unless the employers can show a strong connection to the actual requirements of the job.[44] Employers must show that the practices are justified by "business necessity."[45]

The appropriate justification of the disparate impact standard is widely disputed and raises fundamental questions about the nature of the antidiscrimination principle.[46] The standard can be defended in two different ways.[47] First, it might be seen as a way of ferreting out some kind of illegitimate motive – and might therefore be essentially equivalent, at the level of basic principle, to the disparate treatment standard. Lacking the tools to uncover bad motives, the legal system might ask: Does the manager have a sufficiently neutral justification for adopting a practice that has adverse effects on (say) women? If not, we might suspect that some kind of discriminatory motive is at work.

An alternative defense of the disparate impact standard would not speak of motivation at all.[48] It would insist that if a practice has disproportionate adverse effects on members of traditionally subordinated groups, it should be stuck down, unless it has a strong independent justification.[49] On this view, the motivation of the decision-maker is not relevant. What matters is the elimination of social subordination of certain groups or something like a caste system.[50] The disparate impact standard does not, of course, go nearly that far.[51] But by requiring a strong justification for practices with discriminatory effects, it tends in that direction.

[41] David A. Strauss, *The Law and Economics of Racial Discrimination in Employment: The Case for Numerical Standards*, 79 GEO. L.J. 1619 (1991).

[42] This is a clear implication of Craig v. Boren, 429 US 190 (1976), and JEB v. Alabama ex rel. TB, 511 U.S. 127 (1994). For some complications, see Nguyen v. INS, 533 U.S. 53 (2001). *See* John J. Donohue III, *The Law and Economics of Antidiscrimination Law*, (Nat'l Bureau of Econ. Research, Working Paper No. 11631, 2006), https://www.nber.org/papers/w11631.

[43] The defining decision is Griggs v. Duke Power Co., 401 U.S. 424 (1971). For our purposes, the full intricacies of the doctrine do not require elaboration. See sources cited in note *supra* for discussion.

[44] Griggs v. Duke Power Co., 401 U.S. 424 (1971). The disparate impact standard is now under constitutional scrutiny, but we bracket those issues here. See Richard A. Primus, *Equal Protection and Disparate Impact: Round Three,* 117 HARV. L. REV. 493, 498 (2003).

[45] *See* Christine Jolls, *Antidiscrimination and Accommodation*, 115 HARV. L. REV. 642 (2001); Jake Elijah Struebing, Note, *Reconsidering Disparate Impact Under Title VII: Business Necessity as Risk Management*, 34 YALE L. & POL'Y REV. 499 (2016); Linda Lye, Comment, *Title VII's Tangled Tale: The Erosion and Confusion of Disparate Impact and the Business Necessity Defense*, 19 BERKELEY J. EMP. & LAB. L. 315 (1998).

[46] *See* Strauss, *supra* note; George Rutherglen, *Disparate Impact, Discrimination, and the Essentially Contested Concept of Equality*, 74 FORDHAM L. REV. 2313 (2006).

[47] *See* Pamela Perry, *The Two Faces of Disparate Impact Discrimination*, 59 FORDHAM L. REV. 523 (1991). On the relevant history, *see* Reva Siegel, *The Constitutionalization of Disparate Impact*, 106 CALIF. L. REV. (forthcoming 2019). On the theory, *see* Jack Balkin & Reva Siegel, *The American Civil Rights Tradition*, 58 U. MIAMI L. REV. 9 (2003).

[48] *See* Owen Fiss, *Groups and the Equal Protection Clause*, 5 PHIL. & PUB. AFF. 107 (1976); Cass R. Sunstein, *The Anticaste Principle*, 92 MICH. L. REV. 2410 (1993).

*Fair representation*. Some people, of course, go beyond both disparate treatment and disparate impact. They want members of certain groups to be chosen at socially desirable rates. Call this *the principle of fair representation*.[52] The desire for fair representation may derive from numerous sources. For example, fair representation might have an instrumental value.[53] Having greater numbers of African-Americans in the police force could be important if it improves the functioning of the force, whose relationship to the relevant community might be better if it is not entirely or mostly white.[54] There may also be collateral benefits – including externalities – to such inclusion. Numerous economic examples fit this condition. African-Americans lacking past credit records may not be viable borrowers for lenders, which in turn means they will not ever be able to build a credit record. Even if each individual borrower acts fairly in a local sense, the externality ensures a global unfairness.[55]

The law does not concern itself with fair representation as such, and indeed an effort to pursue that goal, through race-conscious policies, would itself raise serious legal problems under the Constitution and civil rights statutes.[56] Race-conscious affirmative action programs remain constitutional, but only under narrow conditions.[57] Our only claim here is that some people think that fair representation, as such, is a legitimate and important goal.

This discussion thus far should be sufficient to reveal that as a normative matter, the line between antidiscrimination principles and affirmative action is less clear than is often thought.[58] Suppose that an employer is forbidden to take account of customer tastes (even if he deplores them), or that an employer is prohibited from considering statistical reality (even if he deeply wishes it were otherwise). By hypothesis, he is not biased in any way; he is simply trying to succeed in business. Nonetheless, he is required to sacrifice his own economic interests for the sake of attaining broader social goals.[59] In such cases, there is a clear tradeoff. The sometimes-blurry theoretical boundary between anti-discrimination law and affirmative action policy has not posed much of an issue in a world of human decision-making, since the nature of the tradeoffs are hard or even almost impossible to see when so little can be quantified. But this changes when algorithms are introduced, as we discuss below.

## III. ANTI-DISCRIMINATION LAW FOR HUMANS

[49] *See* Ruth Colker, *Anti-Subordination Above All: Sex, Race, and Equal Protection,* 61 N.Y.U. L. REV. 1003 (1986); Owen Fiss, *Groups and the Equal Protection Clause*, 5 PHIL. & PUB. AFF. 107 (1976).
[50] Susan Carle, *A New Look at the History of Title VII Disparate Impact Doctrine*, 63 FLA. L. REV. 251 (2011).
[51] *See* Ian Ayres & Peter Siegelman, *The Q-Word as Red Herring: Why Disparate Impact Liability Does Not Induce Hiring Quotas*, 74 TEX. L. REV. 1487 (1996).
[52] This principle was at the heart of Mobile v. Bolden, 446 U.S. 55 (1980).
[53] *See* Sounman Hong, *How Racial Diversity Makes Police Forces Better*, WASHINGTON POST (Dec. 5, 2017), https://www.washingtonpost.com/news/monkey-cage/wp/2017/12/05/how-racial-diversity-makes-police-forces-better/?utm_term=.7d41f485c23e.
[54] *See* Kathleen Sullivan, *Sins of Discrimination*, 100 HARV. L. REV. 78, 94 (1986).
[55] *See* Cass R. Sunstein, *Why Markets Won't Stop Discrimination*, 8 SOC. PHIL. & POL'Y 22 (1991).
[56] *See* Gratz v. Bollinger, 539 U.S. 244 (2003).
[57] *See id.*
[58] *See* Strauss, *supra* note.
[59] *See* Sunstein, *supra* note.

We will focus throughout on a single category of decisions. Call them "screening" decisions. For example, a manager or team of managers must screen candidates and decide whom to hire. At its essence, decisions in this category are those where:

- We must make a choice about people. Hire or not? Promote or not? Jail or not? Give credit to or not? The decision materially affects people's lives.
- We must use features of people within the relevant class to make this decision. What is their education? What is their past arrest record? How much do they earn? Have they ever defaulted on loans in the past?
- There is some output or outputs the decision-makers say they care about – job performance, loan repayment, flight risk, public safety.
- There is some uncertainty (at the least from the perspective of a third party or a judge) as to which combination of features best predicts the outcome of interest. This uncertainty is both qualitative and quantitative. Does height matter at all for job performance in a physical job? And how much exactly does occupation matter for predicting default given that we already have income data?

In each of these decisions, managers will in effect, implicitly or explicitly, have a screening rule, simple or complex, they are using. We discuss in what follows how the law currently tries to identify and prevent people from engaging in discrimination. But in order to understand how and why we have developed our current legal rules regarding discrimination for humans, it is useful to understand first how human cognition works and exactly what we are worried about.

### A. Human cognition and biases

Remarkably large numbers of Americans admit to discriminatory attitudes when asked. Among white respondents to the General Social Survey (GSS), 28% believe "it's okay to discriminate when selling a home,"[60] 34% say "blacks shouldn't push themselves where they're not wanted," and 45% say "most blacks don't have the motivation or willpower to pull themselves out of poverty."[61]

But the problems run far deeper than even these striking self-reports would suggest. This goes beyond the obvious problem of dissembling – we cannot trust that everyone who discriminates will tell us. The key lesson of a large body of psychology research tells us that people who discriminate often are not aware of it. Much of the time, human cognition is the ultimate "black box," even to ourselves.[62]

---

[60] *See* The Views of White Americans, THE NEW YORK TIMES UPSHOT (2014), https://www.nytimes.com/interactive/2014/04/30/upshot/100000002853274.embedded.html. *See also* Bobo et al., *The Real Record on Racial Attitude*s, in SOCIAL TRENDS IN AMERICAN LIFE: FINDS FROM THE GENERAL SOCIAL SURVEY SINCE 1972 38-83 (Peter V. Marsden ed., 2012).

[61] This is not specific to American race relations; prejudice against out-groups occurs almost anywhere there are people. For example, data from the World Values Survey show:

- 35% of those in Turkey would not want to be neighbors with someone of a different religion
- 41% of South Koreans say they would not wish to be neighbors with an immigrant
- 66% of Russians say they would not wish to be neighbors with someone who was homosexual
- 77% of respondents in India believe that men make for better political leaders than do women

WORLD VALUES SURVEY, *WVS Wave 6*, http://www.worldvaluessurvey.org/WVSDocumentationWV6.jsp?COUNTRY=875 (last visited Jan. 16, 2019).

[62] *See* Wilson, *supra* note.

To understand this point and its implications for discrimination law, it will be useful to say a few words about psychology in general. Dual-processing theories suggest that human cognition consists of two families of cognitive operations: (1) deliberate, conscious thought (what is often called by psychologists "system II thinking"), which is effortful, and (2) rapid, automatic responses, which is not effortful, and of which people may not even be aware ("system I thinking").[63] Because conscious thinking requires effort, we tend to rely on our automatic systems to conserve mental energy.[64] If someone says "two plus two," a number automatically comes to mind in response to this frequently-encountered problem. If you read the word "bread," you might also think "butter."

Automatic responses are not limited to these sorts of small situations; they also extend to behaviors that we would presume are quite mindful, such as how we interact with our social environment. For example, Langer, Blank and Chanowitz asked subjects in their study to make some copies.[65] Just as they were about to do so, a confederate asked to jump in front of them. In some cases, the confederate gave no reason at all ("Excuse me, I have 5 pages. May I use the Xerox machine?"); in others, a reason was given ("Excuse me, I have 5 pages. May I use the Xerox machine, because I'm in a rush?") The third condition had the *veneer* of a reason, but it actually had no informational content ("Excuse me, I have 5 pages. May I use the Xerox machine, because I have to make copies?") People complied at similar rates to both the actual reason and the pseudo-reason (94% and 93%, versus 60% with no reason). What explains this finding? People see a situation they have seen before and their automatic response kicks in, to avoid the effort of processing the rest of this setting. Consistent with this view, when the costs of complying go up – when the confederate wants to copy 20 pages, not just 5 – compliance to the pseudo-reason is similar to what it is in the no-reason condition.

We are often unaware of these automatic responses.[66] Latane and Darley carried out experiments showing that people are less likely to help a stranger in need when there are relatively more people around.[67] They then asked why subjects did not help, "every way we knew how: subtly, directly, tactfully, bluntly. Always we got the same answer. Subjects persistently claimed that their behavior was not influenced by the other people present. This denial occurred in the face of results showing that the presence of others did inhibit helping."[68] Subsequent research has repeatedly confirmed that we often fail to understand why we do what we do.[69] The title of Richard Nisbett and Timothy Wilson's influential essay from decades ago is prescient; in describing our own cognition, we are often "telling more than we can know."[70]

These implicit cognitions can sometimes even be in direct conflict with our conscious thoughts, including with respect to discrimination.[71] The tendency to categorize people and favor "in-groups" and disfavor

[63] The literature is voluminous. For an influential summary and exposition of dual systems models in psychology, *see* Daniel Kahneman, *Thinking Fast and Slow* (2011).Within economics, models of dual systems thinking include Tom Cunningham, *Biases and Implicit Knowledge* (Munich Personal RePEc Archive, Working Paper No. 50292, 2013), and for impulse control, *see* Drew Fudenberg & David Levine, *A Dual-Self Model of Impulse Control*, 96 AM. ECON. REV 1449 (2006).
[64] Daniel Kahneman, *Attention and Effort* (1973), is a classic treatment.
[65] *See* Ellen Langer et al., *The Mindlessness of Ostensibly Thoughtful Action: The Role of 'Placebic' Information in Interpersonal Interaction*, 36(6) J. OF PERSONALITY & SOC. PSYCHOL. 635 (1978).
[66] *See* Geoffrey Cohen, *Party Over Policy*, 85 J. PERS. SOC. PSYCHOL. 808 (2003).
[67] *See* Bibb Latane & John M. Darley, *Bystander Apathy,* 57 AM. SCIENTIST 244 (1969).
[68] *See id.* at 124.
[69] *See, e.g.*, Cohen, *supra* note
[70] *See* Richard E. Nisbett & Timothy DeCamp Wilson, *Telling More Than We Can Know: Verbal Reports on Mental Processes*, 84 PSYCHOL. REV. 231 (1977).
[71] On implicit bias and law, *see* Jolls & Sunstein, *supra* note. On whether implicit attitudes map onto behavior, *see* Hal Arkes & Philip E. Tetlock, ATTRIBUTIONS OF IMPLICIT PREJUDICE, 15 PSYCHOL. INQUIRY 257 (2004); Philip E. Tetlock & Gregory Mitchell, *Implicit Bias and Accountability Systems*, 29 RESEARCH IN ORG. BEHAVIOR 3 (2009).

"out-groups" is ubiquitous,[72] and indeed the automatic system pays particularly close attention to other people's sex, age and race.[73] But the personal characteristics that distinguish groups do not even need to be so obviously distinctive. Consider the famous experiment by Sherif et al. involving two groups of middle-school youth at Robber's Cave State Park in Oklahoma.[74] Over the course of the study, the two groups (the Eagles and the Rattlers) exhibited increasingly negative views about the trustworthiness, integrity, and athletic skill of the other group, even culminating in aggression. This out-group hostility arose even though *the two groups were actually formed by random assignment of a quite homogenous pool of white Protestant boys*.[75] There were no *actual* differences across groups that generated the out-group hostility. All it took was to pit them against each other in a few small competitions.

A major contribution of psychological research has been to show that even implicit biases can be measured.[76] In an early study, Word, Zanna and Cooper asked study subjects – white Princeton undergraduates – to carry out interviews of white and African-Americans confederates.[77] Study subjects were found to sit closer to and spend more time talking with white than African-American interviewees, both measures of how positively inclined one person is toward another. In a follow-up experiment, the subjects now were the ones being interviewed, by a set of confederates randomly assigned to do the interviews either the way whites were interviewed in the first experiment (sitting closer, more time) or how African-Americans were interviewed (further, shorter). Those interviewed as African-Americans had been in the first experiment were rated by observers in this second experiment as having worse interview responses. That is, the interviewer's behavior changes the applicant's responses. Beyond showing that we can measure implicit bias, this study also showed that these biases can create a self-fulfilling prophecy: people are basically creating their own reality.[78]

---

[72] Indeed, when meeting someone new these are typically the most likely things to be remembered about the person, and often the *only* things that are remembered. This automatic encoding may stem from the same basic cognitive processes related to cooperation and paying attention to social coalitions. Some evidence to this effect comes studies that create new social coalitions among study subjects in laboratory conditions that are unrelated to race, which seems to substantially reduce mental encoding of race. *See* Leda Cosmides et al., *Perceptions of Race*, 7 TRENDS IN COGNITIVE SCIENCES 173 (2003).
[73] *See id*.
[74] *See* Muzafer Sherif et al., *Intergroup Conflict and Cooperation: The Robbers Cave Experiment* (1961).
[75] *Id.*
[76] *See* Banaji & Greenwald, *supra* note.
[77] *See* Carl O. Word et al., *The Nonverbal Mediation of Self-Fulfilling Prophecies in Interracial Interaction*, 10 J. EXP. SOC. PSYCHOL. 109 (1974).
[78] A great deal of attention has been devoted to other ways of measuring implicit biases, such as the implicit association test, or IAT. *See* Anthony Greenwald et al., *Measuring Individual Differences in Implicit Cognition: The Implicit Association Test*. 74 J. OF PERSONALITY AND SOCIAL PSYCHOL. 1464 (1998); Marianne Bertrand et al., *Implicit Discrimination*, 95 AM. ECON. REV. PAPERS & PROCEEDINGS 94 (2005). The societal consequences of implicit bias remain difficult to determine, partly because we only have IAT results for convenient (rather than truly representative) samples of respondents. *See* Arkes & Tetlock, *supra* note. But implicit bias seems to be prevalent among those who volunteer to take the tests, and some people argue that it is correlated with actual behavior. For example, counties or metropolitan areas with relatively higher rates of measured implicit bias have been shown to have higher rates of police use of force against blacks, and larger black-white disparities in low birth weight or preterm birth rates *See* Eric Hehman et al., *Disproportionate Use of Lethal Force in Policing Is Associated With Regional Racial Biases of Residents*, 9 SOC. PSYCHOL. & PERSONALITY SCI 393 (2018); Jacob Orchard & Joseph Price, *County-Level Racial Prejudice and the Black-White Gap in Infant Health Outcomes*, 181 SOC. SCI. & MED. 191 (2017). One study administered IATs to managers in a French grocery store chain. They found that the productivity of minority workers was lower when they interacted with managers who had IAT scores indicating more bias. *See* Dylan Glover et al., *Discrimination as a Self-Fulfilling Prophecy: Evidence from French Grocery Stores*, 132 Q.J. OF ECON. 1219 (2017). It turns out managers with higher levels of implicit bias interacted less with minority workers, leading to reduced worker effort. This helped support biased manager stereotypes about minority workers being less productive – another example of a self-fulfilling prophecy.

In sum, with respect to human cognition we are, as suggested by the title of Timothy Wilson's 2004 book, often "strangers to ourselves."[79] In important domains, human behavior often looks quite different from what we think of as the conscious optimization of some clear objective.

Consider in this light the consequence of either conscious or subconscious bias for screening decisions.[80] When managers make a rank-ordered list of job applicants, those who discriminate by (say) gender will put men higher on the list than they deserve to be based on expected productivity. Managers will rank some lower-productivity men above some higher-productivity women because, as in the canonical formulation from Gary Becker, they hire a woman over a man only if her productivity advantage is large enough to compensate the manager for the disutility he gets from hiring a woman.[81] Put differently, the discriminating manager creates a prioritized list of job applicants for hiring that is in some sense *reshuffled*; it is no longer rank-ordered purely by expected job performance. We now turn to the challenge that the law faces in determining when this type of reshuffling has taken place.

### B. The challenge of detecting discrimination

For decades, the challenge of ferreting out illicit motivations has preoccupied analysts of discrimination against protected groups.[82] Indeed, it may be their central preoccupation. We know that racial disparities exist for important outcomes such as income, wealth, credit, schooling, criminal justice, health, health care, and many others.[83] One cause of these disparities involves people's choices and behaviors, including disparate treatment. We know that because people explicitly tell us they discriminate (as in the GSS survey results above), and also from carefully controlled "audit studies." For example, Bertrand and Mullainathan sent fictitious resumes out to a large number of employers in Boston and Chicago that were identical except that half had a white-sounding name and the others had an African-American-sounding name.[84] Resumes with white names received 50% more call-backs from employers who wanted to carry out an interview.[85]

But that evidence for discrimination is *statistical*. Without more, it cannot establish disparate treatment with respect to any *individual* decision. This question about what the former can tell us about the latter is at the heart of the debate about the defining case of *McCleskey vs. Kemp*.[86] An African-American defendant convicted of killing a white police officer and then sentenced to death challenged his sentence by citing statistical evidence that capital punishment was administered at a relatively higher rate in cases with a black offender and a white victim. The Court rejected the challenge and insisted on the need for proof of discrimination in the particular case. It objected that the plaintiff's claim "that these statistics are sufficient proof of discrimination, without regard to the facts of a particular case, would extend to all capital cases in Georgia, at least where the victim was white and the defendant was black."[87]

---

[79] Timothy D. Wilson, *Strangers to Ourselves: Discovering the Adaptive Unconscious* (2004).

[80] As briefly noted in the text, differential treatment on grounds of race, sex, and other factors can arise even when people are not trying to express personal animosity, but merely looking out for the bottom line. Suppose that an employer favors whites over African-Americans, not because he wants to discriminate (he does not), but because his customers like dealing with whites, and he does not want to lose money. Or suppose that a firm relies on a statistical demonstration that on average women leave the workforce more frequently than men do. If a hiring manager cannot tell which *specific* women are more likely to leave, then – if turnover is bad for the bottom line – a profit-maximizing company might use gender as a proxy for the propensity to leave. This is an example of what economists call "statistical discrimination." *See, e.g.,* Kenneth Arrow, *The Theory of Discrimination*, in DISCRIMINATION IN LABOR MARKETS (Orley Ashenfelter & Albert Rees eds., 1973); Edmund Phelps, *The Statistical Theory of Racism and Sexism*, 62 AM.ECON. REV. 659 (1972). Statistical discrimination might be good for the firm, but it penalizes women (including those who are at low risk of leaving the firm themselves). Under the Equal Protection Clause and civil rights laws, it is generally unlawful to discriminate on the basis of customer preferences or to engage in statistical discrimination, even if these are profit-maximizing. See Strauss, *supra* note.

[81] *See* Gary S. Becker, *The Economics of Discrimination* (2nd ed. 1971) (1957).

[82] Charles R. Lawrence III, *The Id, the Ego, and Equal Protection: Reckoning With Unconscious Racism*, 39 STAN. L. REV. 317 (1987); Strauss, *supra* note.

[83] See for example Stanford Center on Poverty and Inequality, *State of the Union, 2017,* https://inequality.stanford.edu/sites/default/files/Pathways_SOTU_2017.pdf

[84] *See* Marianne Bertrand & Sendhil Mullainathan, *Are Emily and Greg More Employable Than Lakisha and Jamal? A Field Experiment on Labor Market Discrimination,* 94 AM. ECON. REV. 991 (2004).

[85] See *id*.

Here is another way to understand the problem. Consider the fact that African-Americans constitute 15% of all college-age young people but that only 6% of all students enrolled at top private universities in the US.[88] This disparity seems more than sufficient to conclude there are important barriers *somewhere* in American life. But for the purposes of the law, we must identify a concrete decision that we believe was affected by discrimination, so that we can name a defendant. Was it the college admissions committee? Or someone else more "upstream" in the process whose decisions contributed to disparities in K-12 learning opportunities? The frequent inscrutability of human decision-making, combined with the relative rarity of explicit archival evidence (memos, verbal statements) of discriminatory motives, makes violations difficult to police.[89]

Without some kind of formal discrimination ("no women need apply"[90]) or a "smoking gun" document, the only other direct way to tell whether someone discriminated in a specific case may be to ask them.[91] Even setting aside the risk they lie, as noted above, they honestly might not even know themselves. Many screening decisions in practice do not involve any sort of formula or guidelines, but instead are largely (or entirely) subjective. A manager may tell us they selected applicants for hiring because they seemed like "good workers." That could mean almost anything. And given the black-box nature of human cognition, even cooperative managers may not be able to explain it. Nor may they be able to articulate what predictor variables for future productivity were used, or why those were chosen over other candidate predictors. And given the possibility of implicit bias, can every manager really say in good faith "I did not pay attention to gender (or race, etc.) in making my decision"? In the analogy to algorithms raised in the introduction, we cannot understand the choices that went into the "trainer" that creates the screening rule the person used. Statistical tests are quite complex.[92]

The black-box nature of the human mind also means that we cannot easily simulate counterfactuals. If hiring managers cannot fully understand why they did what they did, how can even a cooperative manager answer a hypothetical about how he would have proceeded if an applicant had been of a different race or gender? And if it is hard to imagine a counterfactual that involves changing a major salient personal characteristic of that kind, what hope do we have for a counter-factual entailing some incremental change to a different, less salient qualification like educational attainment or years of work experience? Put differently, not only can we not readily understand the "trainer" behind the human's decision, we may also be unable to understand the "screener" that was actually used.

[86] 481 U.S. 279 (1987).
[87] *See id.* at 296.
[88] *See* Jeremy Ashkenas, et al., *Even With Affirmative Action, Blacks and Hispanics Are More Underrepresented at Top Colleges Than 35 Years Ago*, THE NEW YORK TIMES, (Aug. 24 2018), https://www.nytimes.com/interactive/2017/08/24/us/affirmative-action.html.
[89] *See* Carle, *supra* note. An influential treatment is Charles R. Lawrence III, *The Id, the Ego, and Equal Protection: Reckoning With Unconscious Racism*, 39 STAN. L. REV. 317 (1987).
[90] Reed v. Reed, 404 U.S. 71 (1971).
[91] Statistical measures may of course be relevant. *See* Barbara Norris, *A Structural Approach to Evaluation of Multiple Regression Analysis as Used to Prove Employment Discrimination: The Plaintiff's Answer to Defense Attacks of "Missing Factors" and "Pre-Act Discrimination,"* 49 LAW & CONTEMP. PROBLEMS 63, 65.; Kingsley Browne, *Statistical Proof of Discrimination*, 68 WASH. L. REV. 477 (1993).
[92] *See id.*

Of course, courts have developed tools for addressing these challenges, including statistical analysis.[93] If we cannot get a meaningful answer from people about what they did, we might at least try to reconstruct their decision-making by interrogating data, as courts often do.[94] But a data-driven investigation faces its own challenges.[95] For example, it can be difficult for judges to know whether a screening rule treats equally productive candidates the same when they cannot really measure output, but only qualifications. Consider how difficult defining the "output" of a job is even for a relatively simple occupation like, say, retail cashier. Is output the fraction of days worked where the cash register and recorded sales balance out? Or how often the worker shows up on time, not late? Or how often customers come back to the store? Or total sales more generally?

What if we focused on qualifications? With the aid of existing tools,[96] determining whether applicants with the same qualifications are treated the same may be helpful or even sufficient, but it runs into the problem of which qualifications matter. Attempts to focus on those qualifications that are most predictive of output runs into the problem of measuring output. We might be tempted to solve this problem by comparing applicants who are the same on *every* qualification. But this can often lead to a *very* long list. And in many cases there are just not that many people actually hired into a given job by a given firm over a given time period. If we have, say, 10 people hired into a job over the past 5 years but 20 plausibly relevant qualifications, it becomes impossible to tell if a firm really hired someone from an advantaged group over a "similarly qualified" person from a disadvantaged group.

With these questions, we do not mean to suggest that the problem of ferreting out discrimination is insuperable. A great deal of illuminating work casts light on that problem and on potential solutions.[97] The only point is that when human behavior is involved, it can be extremely challenging to ascertain relevant motivations and hence to see whether disparate treatment was at work.

We have seen that in the face of a demonstration of disparate impact, statistical evidence alone is not enough to conclude there was unlawful discrimination; the defendant is given a chance to provide some qualitative evidence of a justifiable, neutral reason for the disparity (such as "business necessity").[98] For example, imagine that the head of a private security firm institutes a rule requiring new employees to be able to run at a specified speed. If these practices have disproportionate adverse effects on some group, such as women, they will be invalidated unless they can show a strong connection to the actual requirements of the job.

Notice this takes us back not only to the difficulty of defining and measuring "output," noted above, but more generally puts judges and jurors in the position of having to make potentially difficult judgments about the best way to carry out some task that is far from their own expertise. The head of the security firm, for example, argues that the ability to run fast is an important part of the job. He notes that there is variability in the running speed of suspects, and variability in the distance of foot-chases, but he believes that a security guard who could run 1.5 miles in 12 minutes should be able to catch the suspect in "most" chases. It is challenging for judges to decide on the merits whether any of the security firm head's assertions are correct. It is no wonder, in these circumstances, that much of the debate turns on an institutional question: how aggressively the legal system should scrutinize those assertions.[99]

---

[93] Kevin Tobia, Note, *Disparate Statistics*, 126 YALE L.J. 2260 (2017); Michael O. Finkelstein, *The Judicial Reception of Multiple Regression Studies in Race and Sex Discrimination Cases*, 80 COLUM. L. REV. 737 (1980).
[94] *See* Segar v. Smith, 738 F.2d 1249 (D.C. Cir. 1984), *cert. denied,* 105 S. Ct. 2357 (1985); *supra* note.
[95] D. James Greiner, *Causal Inference in Civil Rights Litigation,* 122 HARV. L. REV. 533 (2008).
[96] See *supra* note.
[97] See Greiner, *supra* note; Michael O. Finkelstein, *The Judicial Reception of Multiple Regression Studies in Race and Sex Discrimination Cases*, 80 COLUM. L. REV. 737 (1980); Delores A. Conway & Harry V. Roberts, *Regression Analyses in Employment Discrimination Cases*, in STATISTICS AND THE LAW 107 (Morris H. DeGroot et al. eds., 1986); David C. Baldus & James W.L. Cole, *Statistical Proof of Discrimination* (1980).
[98] On the dynamics, see Barocas & Selbst, *supra* note; Linda Lye, Commentary, *Title VII's Tangled Tale: The Erosion and Confusion of Disparate Impact and the Business Necessity Defense*, 19 BERKELEY J. EMP. & LAB. L. 315 (1998); Amy L. Wax, *Disparate Impact Realism*, 53 WM. & MARY L. REV. 621 (2011).

## IV. ALGORITHMS AND HOW THEY WORK

One way to think about the goal of prediction is to overcome a missing information problem. A firm gets applicants for some job. To decide whom to hire, it would like to know each applicant's productivity if he or she were hired. But that information is "missing," in the sense that the applicant has not yet worked for the firm; this is the crux of the question in hiring. Our best guess for what will happen in the future often is what happened in the recent past. So a natural way to fill in (predict) the missing information about the future performance of current job applicants is to look at the performance of "similar" job applicants whom the firm has hired before. The algorithm in this sense is functioning as a data summarizer.

### A. Two Algorithms, Not Just One

We often refer to anything that involves data and the resulting prediction as an "algorithm." But as we have noted, this misses the fact that there are actually two separate "algorithms" in any screening rule application of the sort that we consider here.

One algorithm – what we call the *screener* – simply takes the characteristics of an individual (a job applicant, potential borrower, criminal defendant, etc.) and reports back a prediction of that person's outcome. This prediction for the person's outcome then feeds into a decision (hiring, loan, jail, etc.).

The second algorithm – what we call the *trainer* – is the thing that produces the screening algorithm. This involves (among other things) deciding which past cases to assemble to use in predicting outcomes for some set of current cases, which outcome to predict, and what candidate predictors to consider.

The distinction between these two separate algorithms is often under-appreciated but is actually quite important in practice:

- People often worry that the algorithm could be doing literally anything, including unforeseen things that introduce bias.[100] This concern shows up when, for example, people worry about using machine learning in situations where standard statistical tools like logistic regression have been used to inform predictions for years.[101] Recognizing there are actually two algorithms helps clarify that conditional on the choices that go into constructing the trainer, the screener cannot do "literally anything" – it is mechanically the result of whatever human decisions were made for the trainer.

---

[99] *See* Kevin Tobia, Note, *Disparate Statistics*, 126 YALE L.J. 2260 (2017).
[100] Some of these concerns are noted and addressed in Goel et al., *supra* note.
[101] *See* id.

- The distinction between the screener and trainer is also important for detecting and regulating discrimination, since the way we would carry out audits and monitoring is different for the two algorithms (as discussed further below).

### B. What situations do (and do not) fit this framework

We have in mind situations where the decision we are examining is made many times, so that the training algorithm has enough cases to learn the relationship between the candidate predictors and the specified outcome. This means that algorithms will be better for "micro-prediction" tasks, such as hiring, than for "macro-prediction" tasks, such as who will win a presidential election.[102]

Our framework is relevant to situations where the training algorithm, screening algorithm and training dataset are all fixed, stored objects that can be inspected. In cases where the criterion for the decision is changing at a rapid rate – as in many online learning applications, like Google search or online ad placement – new training observations flow in at a massive rate, and the training algorithm is constantly running to produce new screeners. Things move so quickly in these settings that screeners, trainers and underlying training data samples may not even be stored in a way that would suffice to reconstruct a snapshot of the system at every instant. This means we cannot go back and examine what the training algorithm would have done under different choices. A new legal framework, on the statutory side, might require companies to begin to store the inputs to its algorithms even in settings like this, a point to which we return below.

The settings for which our framework *is* currently relevant affect many people's lives and accounts for a large share of total economic activity each year, settings such as education, hiring, criminal justice, and health care. These decisions do not just have important stakes for society; they are also ones where the United States has had long-standing concerns about the possibility of discrimination.[103]

### C. How algorithms work

We discuss the mechanics of how algorithms work within the context of a concrete example: hiring. The basic steps to constructing a training algorithm involve the following components.[104]

- Collecting a dataset
- Specifying a concrete outcome to be predicted in that data set
- Deciding which candidate predictors to construct and make available to the algorithm to be considered for inclusion in the final statistical model

---

[102] A related reason why machine learning tools are better for "micro-prediction" than "macro-prediction" is that these algorithmic procedures are useful only to the extent to which the new data for which we seek to make predictions are generated by the same underlying data-generating process that produced the data we use to train the algorithm. If the world is fundamentally changing over time, an algorithm built to understand the relationship between X and Y in one state of the world may not carry over to a new state of the world. This assumption of some basic stability to the world is plausible in situations where, for instance, we use data on yesterday's credit card applicants to predict default risk of tomorrow's applicants. But this can break down when we predict low-frequency events like, say, US presidential elections, where the nature of the underlying political dynamics may be changing and we only have one data point every four years to try to discern those changes.

[103] *See, e.g., supra* note.

[104] Textbook treatments of the basic road map for training prediction algorithms can be found in C. Bishop, *Pattern Recognition and Machine Learning*, (2006) and T. Hastie, R. Tibshirani, & J. Friedman, *The Elements of Statistical Learning: Data Mining, Inference, and Prediction* (2[nd] edition, 2009).

- Building a procedure for finding the best predictor, which uses all the other variables to predict the outcome of interest. The result is the screener: we give the screening algorithm a given set of job applicant characteristics, it gives back a prediction of the outcome for that applicant
- Validating the procedure in a "hold out" set – a data set that was not used to train the algorithm on.

The first two steps involve critically important human choices that shape what the algorithm does, and hence the degree to which it might discriminate. For that reason, we discuss them in much greater detail below within the context of regulating algorithmic discrimination, and touch on them only briefly here.

The dataset on past workers can be thought of like a spreadsheet, with each row being someone whom the firm hired in the past. One or more columns would be measures of how each worker performed on the job. The remaining columns might capture application data that can serve as candidate predictors. The human building the algorithm makes decisions about which set of workers to include in the dataset (the rows), and which outcomes and candidate predictors (columns) to invest resources into collecting as well.

The next step is to specify the outcome the algorithm should predict. A human hiring manager might say they are just looking for people who will be "good workers." But for the people building a training algorithm, it is necessary to be more specific. Suppose we are hiring new police officers. When we look at past officers, we can see that some of them are good at making arrests, while others are good at developing rapport with the community. If we specify arrests as the key outcome of interest for the training algorithm, we will wind up ranking highly those current applicants who would be good at making arrests in the future and skew the composition of the department towards people like this. So what outcome is specified, or how different outcomes are weighted together for a "productivity index," matters critically for what the algorithm does and the hiring process that results.

The third step is deciding which candidate variables to make available to the algorithm to be considered for inclusion in the final statistical model. This might include, for example, the construction of new variables from ones that are in the original dataset (like constructing body mass index using height and weight). It also includes decisions about whether the algorithm should not even have the option to consider some candidate predictors in the final model, because for example they suffer from severe measurement error or are disallowed for legal reasons.

The fourth step is the procedure that determines which of the candidate predictors that have been collected should be used in predicting the outcome of interest. Given how important human decisions are in the choice of training data and outcome to predict, it is natural to assume that human judgment plays a critically important role here, too. Indeed, in discussing algorithms that might be used, for instance, to help rank-order police applicants, we find many people ask questions like: Who exactly will be constructing this algorithm? Have they ever been part of the police force? Or lived in a distressed neighborhood where the stakes for picking the right police officers are particularly high? How do we know the algorithm's programmer will not make misguided choices in deciding what factors to use in predicting the outcome?

But this step – deciding which candidate predictors to include in the statistical model, given a full set of candidate predictors and a choice of outcome to predict – is actually the one step in building the training algorithm where the human usually does *not* play so much of a role, or at least plays a role very different from the one that many people imagine. A key feature of machine learning is that the *data themselves* are used to determine what predictors to include in the final prediction model.[105] If being left-handed is unrelated to performance as a police officer, in the training dataset of past police who have been hired we would see that left- and right-handed officers would have similar values of the outcome variable on average – and so the algorithm would choose not to use it as a predictor. If there are instead big differences in average performance of past officers by, say, college degree receipt, the training algorithm would be more likely to select that as a predictor for the final model. The algorithm can only select from among the candidate predictors made available to it by its human designers, but given the choice of training data and outcome, it is the underlying relationships between the variables in the training data that determines which predictors wind up in the screener, and how much weight they get; it is basically just a statistical matter of which variables are most correlated with the outcome.

[105] *Id.*

This helps us see that it is *not* the case the training algorithm could do "literally anything" at this step. Another way to think about what the algorithm is doing at this step is to use the candidate predictors (job application information) to group past workers together based on who has similar job performance. We then essentially predict the future performance of current applicants as being something like the average performance of past workers in the same group.

This type of "grouping" is at the heart of prediction, whether that prediction is done by a human being or by an algorithm. The whole idea of prediction is sometimes viewed as objectionable because it has the flavor of profiling. But if we truly believed that every job applicant was fundamentally different from everyone else, then there is literally nothing we could say about how they would perform on the job unless we actually hired them and observed what happens. If we really thought people were so idiosyncratic that we could never learn anything about them from observing what happened with anyone else, there would be no way to develop a "Joseph Michael Smith, DOB 1/1/1980" predictor – because there is only one such person. If we gave up on the prediction, on what basis would hiring be done at all? Moreover, as have noted above, the designers of the algorithm have the ability to specify, and thereby restrict, the set of candidate predictors available for the procedure to consider. The alternative to machine prediction is some combination of much-less-accurate human prediction, and other subjective human decision-making of the sort that is fraught with risk of bias.

A final question we might have is: How do we even know that these algorithms actually work? This is why the last step in the process is so critically important – we must validate the algorithm's performance in a new dataset that the algorithm has not seen yet during the algorithm training process. Consider the Netflix Challenge, where research teams competed to develop the most accurate possible prediction model in response to a publicly available dataset.[106] The only reason the Netflix prize competition worked was that the company made one dataset available to research teams to develop their prediction models on, but then evaluated their performance on a separate dataset that was not made public. Imagine Netflix had made the hold-out set public. We would be worried that if the competing teams had access to the hold-out set, they would simply try a vast number of candidate algorithms until they stumbled across one that happened to fit the idiosyncrasies of that particular dataset well, rather than uncover a generalizable way in which nature works that would show up consistently across different datasets. In many applications we do not have the same level of control over the hold-out set, and so have no choice but to trust the algorithm builder not to have looked at the hold out set during the algorithm construction stage.

This description of how the training algorithm works also makes clear why the choice of screening algorithm is in some sense the inevitable byproduct of statistical principles once the choice of outcome, candidate predictors, and training dataset have been made. The real fear with algorithms is not what happens inside the algorithmic "black box." The training algorithm is in some sense the *opposite* of a black box; the algorithm summarizes the data according to the data given to it and the outcome that is specified. It is those human choices where the potential for problems, including discrimination, really arise.

[106] For an outline, see RM Bell and Y. Koren, *Lessons from the Netflix Prize Challenge*, 9 SIGKDD EXPLORATIONS 95 (2007).

### D. What the algorithm *cannot* do

The training algorithm is better at prediction than it is at causal inference.[107] It is designed to identify a collection of predictors that help predict the outcome as accurately as possible among a new set of observations. Suppose there are multiple candidate predictors in a dataset that are highly correlated with one another – for example, given the high levels of racial segregation that persist in the US, race might be highly correlated with both neighborhood of residence and high school attended. An algorithm will pick whatever subset of them happen to be most useful for prediction purposes, but may not include *all* of the correlated variables. One implication is that if we see a predictor included in the model, we cannot tell whether it is that predictor or some other correlated predictor that is actually causally affecting the outcome.[108] Another implication is that if we see an algorithm that does not include a protected personal attribute like race in the final model, that does not mean that a correlated proxy for race is not playing a role. It is worth underlining this point: An algorithm that is formally blind to race or sex might be using a correlated proxy. Whether that is a problem, for legal purposes, depends on the governing legal standard. It is more obviously relevant to a disparate impact claim than to a disparate treatment claim.

There is a different way in which the algorithm is better for prediction than causation. Suppose we are trying to predict performance on the job at some firm that has created a hostile workplace environment for women. Because of that hostile environment, we may see lower levels of productivity on average for women the firm has hired in the past compared to men.[109] Changing the firm's workplace to make it more accepting and accommodating for women may well improve the average productivity we see among future female hires,[110] but the algorithm cannot speak to this possibility. The algorithm is a tool designed for prediction (how do women do at this type of workplace?), not causal inference (what happens if we change the work environment?).

Another important limitation is that we can train the algorithm only using observations for which we observe the outcome. We cannot observe productivity on the job at some firm for, say, women if women never apply to that place to work (as they might not if the firm has a hostile workplace). We also cannot observe productivity on the job for women at some firm if the firm never hires women. This limitation implies that past choices have a powerful impact on prediction, by determining for whom we have outcome values. We return to this point in the next section when discussing the important role played by the human choice of training data to give to the algorithm.

Recall that the training algorithm can only optimize whatever outcome, candidate predictors and training data are given to it. The flip side is that this is *all* the algorithm does. It has no ulterior motives or hidden agenda. And much of what it does is transparent to us.

---

[107] *See* Jon Kleinberg et al., *Prediction Policy Problems*, 105 AM. ECON. REV. 491 (2015).
[108] *Id.*
[109] For relevant evidence and discussion, *see* Amna Anjun et al., *An Empirical Study Analyzing Job Productivity in Toxic Workplace Environments,* 15 INT. J. ENVIRON. RES. PUBLIC HEALTH 1035 (2018).
[110] *Id.*

That is unlike with the human being. Eye-tracking studies of how HR personnel screen resumes tell us that the average hiring manager spends an average of merely six seconds looking at each resume, with 80% of the time spent on just six items.[111] How people decide which factors to look at, or how to weight them together, remains largely a mystery.

## V. WHERE DISCRIMINATION CAN (AND CANNOT) ARISE WITH THE ALGORITHM

One concern with algorithms is that they may seem technocratic and dispassionate in a way that creates the veneer of unimpeachable objectivity. But as we have noted in the previous section, given the important role that human decisions play in the construction of algorithms, it would be irresponsible – even dangerous – to confuse "data-driven" with "nondiscriminatory," "unbiased," or "objective."[112] In what follows, we outline how discrimination in the human choices about outcomes, candidate predictors, and training procedures can infect the resulting algorithm; as we demonstrate in the Appendix to the paper, the overall level of discrimination can be fully decomposed into contributions from these three components. We also discuss ways in which we might worry that algorithms might be discriminating, but are not really; these sorts of misconceptions can lead to restrictions that are well-intentioned but ultimately counter-productive.

It may be useful, by way of preface, to distinguish among four kinds of problems. *First*, the algorithm might be engaging in disparate treatment – as, for example, if it considers race or gender and disadvantaged protected groups (perhaps because racial or gender characteristics turned out to be relevant to the prediction problem it is attempting to solve). *Second*, the algorithm might be producing a disparate impact – as, for example, if it considers some factor whose usage disproportionately burdens members of protected groups (say, certain kinds of test scores). *Third*, the algorithm might be considering some factor that is itself a product of past discrimination (as might be true, for example, of credit scores or arrest records). The law does not concern itself with this problem except insofar as it reflects disparate treatment or can make out a case of disparate impact in the relevant domain, but it might nonetheless be counted as a problem. *Fourth*, the algorithm might produce an imbalance of a kind that many people find objectionable. The law is not concerned with this problem, at least not by itself,[113] but many people would be concerned to see that an algorithm ends up giving (for example) some benefit to women far less than to men.

### A. Where discrimination *can* arise with algorithms

### 1. Choice of outcome

We noted above that what outcome to predict, or how to weight together different outcomes, is one of the critical choices that must be made in building the training algorithm. The choice of outcome is non-trivial even in the simplest of examples. Consider for example a data-entry firm, where employee output seems easily quantified through the amount of data entered. Even here, a firm might choose to measure output through number of hours at work – arguing that an unused computer is a wasted resource. Such a choice is non-trivial since home-life differences may lead to gender differences in hours at work.[114] Gender differences that are bigger for some outcomes than others open the door to intentional bias by the algorithm constructor. It is possible that the case is one of disparate treatment. It might instead involve disparate impact.

[111] The six items are: Name; current title and company; previous title and company; previous position start and end dates; current position start and end dates; and education. *See* THE LADDERS, *Keeping an Eye on Recruiter Behavior* (2012), https://cdn.theladders.net/static/images/basicSite/pdfs/TheLadders-EyeTracking-StudyC2.pdf.

[112] *See* Goel et al., *supra* note.

[113] Mobile v. Bolden, 446 U.S. 55 (1980).

Disparate impact could easily be introduced inadvertently. Perhaps the most important way in which this could arise is if the training algorithm is asked to predict a *human judgment*, rather than an external measure of the underlying outcome of interest itself. Suppose for instance that instead of predicting how productive a worker is on the job in the future, an algorithm is asked to make screening decisions based on the people who were hired in the past. Any human bias in past hiring decisions will then be baked into the algorithm itself. This is a form of biased data that the algorithm builder may or may not be aware of, but is quite pervasive – arising, for example, with manager assessments of worker performance, or with customer ratings of workers. Use of such data could well give rise to disparate impact and, on certain assumptions, could constitute disparate treatment as well.

### 2. Choice of predictors to collect, construct, and give to the trainer to consider

Both disparate treatment and disparate impact can also be introduced through decisions about what candidate predictors to collect, construct, and give to the training algorithm to consider for possible inclusion in the final statistical model. For example, an employer that is screening applicant resumes might believe that college quality could matter in predicting worker performance. It might then invest time to assemble US News rankings of four-year colleges (which whites attend at relatively higher rates) without investing the same effort in measuring rankings of two-year colleges or for-profit universities (which black students attend at relatively higher rates).[115]

The problem can arise not just when the firm collects information that is more favorable to one group than another, but also when the firm simply collects too little information overall. For instance, imagine that a firm collects information about where applicants attended high school, but does not collect any information about how someone did in high school. High schools continue to be highly segregated on the basis of race in the United States, and minority students continue to attend high schools of lower average quality than those attended by whites.[116] If academic performance is positively related to performance on the job, an algorithm trained using just data on where someone went to high school might have fewer minorities ranked highly on the hiring list than would one that also collected information about individual student performance. Additional information on academic outcomes allows the algorithm to identify those students who manage to do well in high school despite having attended a disadvantaged, under-resourced school.

The human algorithm builder might also choose to collect candidate predictors that are human judgments, rather than objective measures of reality, and these can lead to difficulties in much the same way that using human judgments as outcome variables can cause problems. Consider the problem of predicting who will do well in college, as a way to inform admissions decisions. Many, or even most, selective schools rely on teacher recommendation letters as part of the admissions process. But previous research suggests that teachers have biases that run along both gender and race lines.[117] Using teacher recommendations to predict student performance may bake teacher bias into the university's prediction of the college performance of applicants.

[114] As another example, among financial analysts previous research suggests men on average may have slightly more accurate earnings forecasts than women, but women have better overall professional reputations. *See* Clifton Green et al., *Gender and Job Performance: Evidence from Wall Street*, 65 FIN. ANALYSTS J. 6 (2009).

[115] Almanac 2018, *Racial and Ethnic Distribution of Students Among Higher Education Sectors, Fall 2016,* THE CHRONICLE OF HIGHER EDUCATION, (Aug. 19, 2018), https://www.chronicle.com/article/Distribution-of-Students-Among/244069.

[116] *See* Linda Darling-Hammond, *Unequal Opportunity: Race and Education,* BROOKINGS, (March 1, 2018), https://www.brookings.edu/articles/unequal-opportunity-race-and-education/.

### 3. Choice of training procedure

Finally, once the designers of an algorithm have chosen an outcome measure and a set of input variables, they need to run a training procedure based on past data in order to produce the screener. Discrimination can easily be introduced in this phase, deliberately or unintentionally. Perhaps most simply, the designers could under-optimize in the training phase and produce a screener that helps some groups and hurts others while only poorly approximating the intended outcome measure. Such under-optimization, which could amount to disparate treatment, should be detectable by a separate party auditing the firm's procedures if they are provided with the choices of outcome, input variables, and training data, since they could then run their own version of the training procedure and see if they obtain a significantly more accurate approximation to the intended outcome measure.

The more subtle, complex routes for discrimination in the training procedure tend to arise through the choice of the training dataset used to produce the screening. At the most basic level, humans may in principle deliberately introduce discrimination into the algorithm through their construction of the training data. Suppose a firm wanted to establish a "bad track record" for minority workers to avoid hiring them in the future. It could intentionally select the *least* productive minority workers from its applicant pool for some period of time.[118] Using the data on how these workers performed on the job to train an algorithm would then predict a disparity in worker productivity even if minority and white applicants in the larger population are equally productive.

Beyond these relatively extreme examples, there are many ways in which the choice of training sample could also *inadvertently* produce some kind of problem, possibly in the form of disparate impact. Suppose that we are trying to predict college success to inform admission decisions. We use as training data a set of records from a university that has a campus climate hostile to female students. This would depress observed academic performance of women and hence lead to systematic disparities in predicted performance of future college applicants from these groups.[119]

A major problem that occurs repeatedly, and is often overlooked, is that we can only predict outcomes using training observations about those for whom we observe the outcome. Suppose we have a firm to which women simply do not apply (perhaps because it has a hostile workplace environment toward them), or the firm just does not hire many women (for one reason or another). This means we sometimes have to rely on very small samples to estimate predictions for some groups, and will predict outcomes for the majority group more accurately.

---

[117] *See* Thomas S. Dee, *Teachers, Race and Student Achievement in a Randomized Experiment,* 86 THE REV. OF ECON. & STAT. 195 (2004); John J. Donohue III, *The Law and Economics of Antidiscrimination Law*, (Nat'l Bureau of Econ. Research, Working Paper No. 11631, 2006); Victor Lavy & Edith Sand, *On The Origins of Gender Human Capital Gaps: Short and Long Term Consequences Of Teachers' Stereotypical Biases* (Nat'l Bureau of Econ. Research, Working Paper No. 20909, 2015).

[118] *See* Dwork et al., *supra* note.

[119] Of course, if we had a training sample from multiple college campuses and collected some information about how minority or female students rated their experience at different places, the algorithm would have an opportunity to learn that disadvantaged groups perform less well at universities that have a hostile campus climate against them.

A subtler problem arises from the fact that our datasets for applications like hiring are filtered through past decisions (who applies, who gets hired), and the humans who make those decisions often have access to information that is not captured in any dataset (for example, what happens during an in-person interview). Suppose we have a man in charge of hiring for a firm who feels threatened by women who are "too competent." As a result, he hires women only if they had below-average interviews. If interview performance is correlated with performance on the job, an algorithm trained using data just on those women who get hired will understate the performance of women relative to men for a given set of observable applicant characteristics.

This critically important problem – what we called the *selective labels problem* – has received relatively little systematic investigation in both the scientific literature and practical applications of algorithms to social science or policy settings.[120] The solution to this problem is likely to require combining insights from computer science with the sort of "natural experiment" research designs developed in social science.[121]

Note that if we took any of the training datasets described above and asked *humans* to learn about the relationship between the outcome and the candidate predictors, they would likely form impressions that are mistaken in the same way that the algorithm would. Put differently, the discrimination that gets introduced into the training algorithm from biased training samples is not indicative of a problem with what the *algorithm* is doing, but rather with the skewed data itself.

### B. Where discrimination is unlikely to originate with algorithms

Having now seen the three fundamental ways in which discrimination can arise in the design of the algorithm, we return to the converse point (developed more formally in the Appendix): any discrimination in the algorithmic screening can be fully decomposed into contributions from these three sources. This means, in particular, that a number of other dimensions of the algorithm's behavior do not represent avenues for potential discrimination to originate. It is useful to list some of these explicitly, as they form the basis for common misconceptions about algorithmic bias.

#### 1. Discriminatory selection of what candidate predictors to include in prediction function

Observers often worry that literally anything could happen within the machine learning black box, which could in turn lead to discrimination.[122] This concern can be seen, for instance, in objections to the use of machine learning algorithms for prediction tasks that have long relied on standard statistical tools like logistic regression.[123] But if the algorithm designer has fixed the outcome, the set of available input variables, and the training procedure, then standard machine learning algorithms will select the subset of input variables that optimize the resulting estimate of the outcome measure. There should be a genuine concern whether the designer may have omitted a crucial variable from the available input that the training algorithm considers for inclusion in the screener, or included variables whose use constitutes discrimination as described above, but less of a concern that the training algorithm will include the wrong variables from the set of available options when it produces the final screener. Not coincidentally, this step of choosing the optimal subset from among the available input variables is a part of the process that is data-driven, rather than closely programmed by a human.

[120] *See* Kleinberg et al., *supra* 7.
[121] For an excellent overview of the natural experiment style of data analysis, *see, e.g.,* Joshua Angrist & Jörn-Steffen Pischke, *Mostly Harmless Econometrics: An Empiricist's Companion* (2009). For an example of how this type of research design can be combined with machine learning predictions to overcome the selective labels problem, *see* Kleinberg et al., *supra* 7.
[122] *See, e.g.,* Yavar Bathaee*, The Artificial Intelligence Black Box and The Failure of Intent and Causation,* 31 HARV. J. L. TECH. 889, 920 (2018).
[123] For relevant discussion, see Michael L. Rich, *Machine Learning, Automated Suspicion Algorithms, and the Fourth Amendment,* 164 U. PA. L. REV. 871 (2016); David Lehr and Paul Ohm, *Playing with the Data: What Legal Scholars Should Learn About Machine Learning,* 51 U. C. DAVIS L. REV. 653 (2017).

Of course, the algorithm designer could always choose to introduce discrimination manually at this stage by deviating from the standard pipeline for building the screener. Nothing that we have said rules out the possibility that a discriminatory algorithm builder might try to hide something deep in the code that intentionally reduces the predicted level of future productivity for every female job applicant. Or, as noted above, the algorithm builder might under-invest in the machine learning engineering that is required to derive the most accurate prediction function possible with a given dataset. The fact that machine learning requires skill and effort to optimize is what gives rise to competitions like the Netflix Challenge described above.[124]

But if the outcome, the input variables, and the training data are made available, there is a natural approach for detecting this type of deliberate discrimination, given that the optimization of the training algorithm is a data-driven process that should be determined by the underlying statistical relationship in the data between the outcome and the candidate predictors. Specifically, third parties could run their own training procedure from the same starting point and determine whether they are able significantly to improve on the firm's predictions. Malfeasance here is relatively feasible to detect, and hence it should be correspondingly feasible to deter.

### 2. The screener algorithm

Recall that the design of a prediction algorithm for a screening problem involves two algorithms: the trainer and the screener. Our discussion thus far has been about the trainer, and the multiple ways in which discrimination can be introduced through the choice of outcome, input variables, and training procedure. Once we have accounted for all these components, the screener itself has essentially been determined: we should think of it as the mechanical result of applying a training procedure to an outcome, a set of input variables, and a set of training data. By itself, it cannot add more bias than what has appeared in the construction procedure that led to it, via the routes discussed above.

With this in mind, consider what disparate treatment looks like. For hiring, an HR manager might do something like rank-order a less-productive white applicant over a more-productive minority applicant. But with a training and a screening algorithm, if we have accounted for the possible sources of discrimination in the human's choices for the training algorithm (outcome, candidate predictors, and training procedure), then there is essentially no room for the screening algorithm to discriminate beyond what has already been accounted for in its construction.

So much, then, for disparate treatment by the screening algorithm. In the world of algorithms, that form of discrimination should be detectable as a deviation from the expressed specification for the algorithm's behavior, so long as we have the ability to audit the training algorithm's design and interact with the screening algorithm itself.

Moreover, if we have access to the screening algorithm, then we need never be surprised by its decisions. We can simulate screening decisions for individual cases or whole populations in a way that would be impossible to imagine in scenarios where a human was performing the screening. For example, we can ask about average rates of acceptance by the screener over subsets of the population. We can also simulate counterfactuals. Given access to the operation of the screener, it is trivial to ask "With this screening rule, would this candidate have been hired if characteristic X had changed?" (For example, if certain applicants had a high school diploma instead of being high school dropouts, would their predicted outcome change enough to move them up enough in a firm's rank-ordered hiring priority list to have been offered a job?)

[124] See Andrey Feuerverger, *Statistical Significance of the Netflix Challenge,* 27 STATISTICAL SCIENCE 202 (2012); Bell & Koren, *supra* note.

In all of these respects, screening algorithms are altogether different from human screeners. Far from being a "black box," they are far more transparent than humans.

### 3. Group differences in the raw data and predicted outcomes

Concerns about biased data, potentially making out either disparate treatment or disparate impact, are very real, as we have discussed in detail above. But systematic differences across groups in the data – that is, in the distributions of the candidate predictors, or in the outcome variable to be predicted – are not by themselves proof that use of the data is discriminatory in any legally relevant sense.[125]

We live in a world in which reality itself systematically differs across groups. For example, we know that women shoulder greater child-care burdens than do men.[126] When we see in some dataset that women earn less than men, it is not necessarily the case that the data are biased in the sense that they systematically *mismeasure* reality for some groups. The underlying reality itself may be a product of discrimination of some kind. Or consider a dataset that shows differences in average arrest rates or test scores for minorities and whites.[127] We know that the poverty rate in the US is twice as high for blacks than for whites;[128] among the poor, blacks are twice as likely as whites to live in high-poverty (>40%) neighborhoods.[129] The average net worth of black households is only about one-seventh that of whites.[130] Given the important role that family and community circumstances play in shaping people's opportunities growing up, it would be remarkable if these large differences in socio-economic circumstances did not lead to average differences in children's outcomes across groups. If the underlying predictors and/or outcomes differ systematically across groups, any well-calibrated algorithm will yield group differences in predictions.

It is possible that we might wish to adjust our algorithmic predictions or subsequent decisions in ways that help address the underlying social problem. This is not the task of eliminating bias that has been introduced by the algorithm, but the task of using the algorithm to address bias and structural disadvantage in the world. We return to this issue below.

---

[125] For a useful application, see David Williams & Selina A. Mohammed, *Discrimination and Racial Disparities in Health: Evidence and Needed Research,* 32 J. BEHAV. MED. 20 (2009).
[126] *See* Henrik Kleven et al., *Children and Gender Inequality: Evidence from Denmark*, (Nat'l Bureau of Econ. Research, Working Paper No. 24219, 2018).
[127] *See, e.g.,* Lauren Nichol Gase et al., *Understanding Racial and Ethnic Disparities in Arrest: The Role of Individual, Home, School, and Community Characteristics,* 8 RACE SOC. PROB. (2016); Christopher Jencks and Meredith Phillips, *The Black-White Test Score Gap: An Introduction* (2013), available at https://www.brookings.edu/wp-content/uploads/2013/01/9780815746096_chapter1.pdf.
[128] *See* Pew Research Center, *On Views of Race and Inequality, Blacks and Whites are Worlds Apart*, PEW RESEARCH CENTER: SOCIAL AND DEMOGRAPHIC TRENDS (Jun. 27, 2016), http://www.pewsocialtrends.org/2016/06/27/1-demographic-trends-and-economic-well-being/.
[129] *See* Elizabeth Kneebone & Natalie Holmes, *U.S. Concentrated Poverty in the Wake of the Great Recession,* BROOKINGS INSTITUTE (March 31, 2016) https://www.brookings.edu/research/u-s-concentrated-poverty-in-the-wake-of-the-great-recession/.
[130] *See* Edward N. Wolff, *The decline of African-American and Hispanic Wealth since the Great Recession*, (Nat'l Bureau of Econ. Research, Working Paper No. 25198, 2018).

## C. What questions we need to ask

With this framework for analyzing algorithmic discrimination in place, we now ask how to ferret out discrimination in a new world that involves algorithms in the loop. We discuss these issues within the context of a concrete example (hiring), and contrast what we want out of a new regulatory framework and how this differs from our existing legal framework.

For orientation, it is useful to situate the standard legal analysis (without algorithms) in this context. Suppose a woman applies to a small business for a job as a salesperson. Her application is rejected. She sues the firm for gender discrimination. For disparate treatment and disparate impact cases, the analysis is straightforward, at least in broad outcome. Recall that disparate treatment might be shown if the employer uses some explicit or implicit rule that treats female applicants less favorably because they are female.[131] If no such rule is in place, the applicant might try to uncover discriminatory motive; that might be difficult.[132] Recall too that disparate impact might be shown if the employer uses some rule (for example, a height and weight requirement) that treats women worse than men. If so, the rule must be justified in some way.[133]

If an algorithm is involved, the plaintiff's lawyers might again seek to show disparate treatment (if, for example, sex is explicitly used as a factor) or disparate impact (if some factor is used that disproportionately harms women). In addition, a key question that the plaintiff's lawyers would want to answer is this: Why did the firm (or whoever built the algorithm for the firm) choose the outcome that it did? One reason to ask that question is to establish disparate treatment.

We have seen that asking this sort of direct question in a situation of exclusively human decision-making is often a challenging exercise. If the algorithm is directed to take account of sex, the issue will be more tractable (with one qualification to which we turn below). And with a standard prediction algorithm in the decision loop, whatever person builds the algorithm will have had no choice but to specify what outcome is being predicted – the training algorithm cannot be built without an outcome at least implicitly specified in the process.

Moreover, the nature of how self-incrimination arises has changed. People are ordinarily reluctant to admit that they discriminated, not least because it subjects them to legal liability. But if the algorithm uses gender (or race, or sex) – even if it is merely directed to predict some outcome – disparate treatment will be more readily apparent. Or consider the response options for someone who has intentionally selected a specific, facially nondiscriminatory outcome to predict in order to discriminate. The firm that chose, say, on-time attendance rate as the outcome to predict might be tempted to say it was chosen because that is what the firm values most in its workers. That response, however, is now hemmed in by the paper trail that has accumulated over the years of the firm's operations that document what the firm has focused on when doing annual performance reviews and making promotions. The "smoking gun" comes not from someone confessing discrimination, but (much easier to obtain) someone making a claim about the selection of an outcome that transparently conflicts with everything else the firm has been doing.

---

[131] *See* McDonnell Douglas Corp. v. Green, 411 U.S. 792 (1973); Watson v. Fort Worth Bank & Trust, 487 U.S. 977 (1988).
[132] See Michael O. Finkelstein, *The Judicial Reception of Multiple Regression Studies in Race and Sex Discrimination Cases*, 80 COLUM. L. REV. 737 (1980).
[133] See *supra* note.

Whether the claim is based on disparate treatment or disparate impact, we might also want the plaintiffs to be able to compare the various choices the firm made in constructing its training algorithm to different relevant internal or external benchmarks, based on the three categories of discriminatory components discussed above:

- How do the hiring disparities that result from the outcome the firm actually chose compare to the disparities that would have arisen if the firm had chosen other candidate outcomes for which the firm has data available? (This question may be relevant to disparate impact.)
- How do the firm's choices of candidate predictors to collect and construct compare to what other firms in the same industry select? (This question may be relevant both to disparate treatment and to disparate impact.) The motivation for this type of industry benchmark comes from evidence that while discrimination remains an important problem, not all firms discriminate.[134]
- How does the training sample the firm used compare to the larger universe of past applicants and hired workers at the firm? How does it compare to the overall population? How do the statistical relationships between the outcome and predictors in the training sample compare to what we have seen in other datasets? (These questions may be relevant both to disparate treatment and to disparate impact.)

None of these comparisons by themselves would constitute sufficient proof of discrimination by the firm. We need to assess the answers in view of the standards for disparate treatment and disparate impact. But if a firm's choice of outcome, candidate predictors or training sample is an outlier relative to the relevant benchmark, we have at least suggestive evidence of disparate treatment. Or disparate impact might be involved. If the firm has a disparity in its hiring outcomes based on (say) choice of predictors, the burden might shift to the firm to justify the decisions it made in constructing the training algorithm.

Notice there can be some counter-intuitive results from this type of benchmarking. Suppose we have a firm that proudly announces it will no longer collect information on whether applicants have a prior criminal record. We compare that to the industry standard and find a similar firm in the same industry and local labor market that proudly announces they invest substantial resources in collecting information about prior criminal records that is as detailed and extensive as possible. Which firm do we imagine will be more likely to hire African-American applicants? A growing body of research in economics suggests that the picture is complicated, in that suppressing information of one type (criminal record) can incentivize decision-makers to turn to other forms of information (race) in ways that may be overtly discriminatory.[135]

---

[134] In the canonical economic model of discrimination from Becker, *supra* note, minority workers – who are a modest share of the workforce in most local labor markets – will wind up seeking out and sorting into the least prejudiced set of employers. *See* Becker, *supra* note 24. So if employers do actually vary in their level of prejudice, and workers sort in the way that Becker's model predicts, then the racial wage gap should be determined by the level of prejudice of the marginal employer. One paper provided some empirical evidence to support this prediction, showing that the prevalence of racial prejudice within a state or region (as measured by survey data from the General Social Survey) is systematically related to the size of racial wage gaps. *See* Kerwin Charles & Jonathan Guryan, *Prejudice and Wages: An Empirical Assessment of Becker's The Economics of Discrimination*, 116 J. POL. ECON. 773 (2008).

[135] *See, e.g.*, Stephen Raphael, *Improving Employment Prospects for Former Prison Inmates: Challenges and Policy,* in CONTROLLING CRIME: STRATEGIES AND TRADEOFFS, (Philip J. Cook et al., eds., 2011) for studies that relate the propensity of firms to check criminal backgrounds with their willingness to hire black applicants, and for studies showing that implementation of a "ban the box" policy that restricts employers from asking about prior record seems to reduce their willingness to hire black applicants, see Amanda Agan & Sonja Starr, *Ban the Box, Criminal Records, and Racial Discrimination: A Field Experiment*, 133 Q. J. ECON 191 (2018); and Jennifer Doleac & Benjamin Hansen, *Does 'Ban The*

We can also run simulated data through the training algorithm, where we know exactly what we should expect in the resulting screening rule, to compare the screening rule that results with what we would expect. For example, we might simulate a dataset where the average level of productivity for black job applicants is higher than for whites. If the screener that results from a given training algorithm gives us back a rank-ordered list of job applicants that nonetheless still hires white applicants at far higher rates than blacks, we would suspect there has been something hidden inside the training algorithm that leads to a discriminatory result, not a truly optimized algorithm.

## D. Problems of Proof: Human Beings vs. Algorithms

Some of these arguments might seem difficult to assess in the abstract. A few stylized cases might therefore be useful by way of clarification:

1. A firm hires salespeople. It favors employees who will maximize sales. It ends up giving a preference to white applicants for one reason: it has found that the firm's customers are more likely to buy from white salespeople.
2. A firm hires managers. Other things being equal, it seeks managers who will stay on the job for a long time. In its experience, women are more likely to leave after a short period – under five years – than men . For that reason, the firm gives a preference to male applicants.
3. A state government is hiring entry-level budget analysts. It gives a preference to applicants from the most prestigious colleges and universities, because those applicants have done best in the past. The preference has a disproportionate adverse effect on African-American applicants.
4. A firm hires security guards. It believes, on the basis of its experience, that security guards with more than thirty years of experience do less well in the job. It therefore gives preference to applicants who have had fewer than thirty years of experience. The result is a disparate impact on older people.

In all four cases, the legal analysis reasonably straightforward. In cases (1) and (2), there is unlawful discrimination. Case (1) involves a form of taste-based discrimination, which is unquestionably a form of disparate treatment, even if the firm is acting rationally.[136] Case (2) involves statistical discrimination, which is also a form of disparate treatment. In both cases, the result is clear.[137] In cases (3) and (4), we need to identify the governing legal standard. Suppose that disparate treatment is forbidden and that if a disparate impact is shown, a substantial business justification is necessary. If so, there would be a legitimate legal challenge in both (3) and (4), not because of disparate treatment, but because the disparate impact would trigger a burden of justification.

While the legal analysis is reasonably straightforward, litigation might not be. For reasons we have discussed, the problem of proof might be formidable in cases (1) and (2): how are we to establish that race and gender were explicitly considered? Cases (3) and (4) might turn out to be similar. If documentary evidence is not available to prove the relevant preferences, a plaintiff might have to make some kind of statistical demonstration. That might be challenging.[138]

---

*Box' Help Or Hurt Low-Skilled Workers? Statistical Discrimination and Employment Outcomes When Criminal Histories Are Hidden* (Nat'l Bureau of Econ. Research, Working Paper No. 22469, 2016).

[136] See *supra* note.

[137] See *supra* note.

[138] See *supra* note.

Now let us consider versions of these cases where an algorithm is used.

5. A firm uses an algorithm to hire salespeople. The algorithm is designed to favor employees who will maximize sales. It ends up giving a preference to white applicants for one reason: the firm's customers are more likely to buy from white salespeople.
6. A firm uses an algorithm to hire managers. Other things being equal, the algorithm seeks managers who will stay on the job for a long time. Relevant data show that women are more likely to leave after a short period – under five years – than men are. For that reason, the algorithm gives a preference to male applicants.
7. A state government uses an algorithm to screen entry-level budget analysts. The algorithm gives a preference to applicants from the most prestigious colleges and universities, because those applicants have done best in the past. The preference has a disproportionate adverse effect on African-American applicants.
8. A firm uses an algorithm to hire security guards. Relevant data show that security guards with more than thirty years of experience do less well in the job. The algorithm therefore gives preference to applicants who have had fewer than thirty years of experience. The result is a disparate impact on older people.

With respect to the underlying law, cases (5) through (8) are the same as cases (1) through (4). In principle, they should not be treated differently.

The real question is whether litigation would be different and whether the plaintiff would find it easier or harder to solve the problem of proof. Finding out in cases (5) and (6) whether the algorithm used an illicit factor (such as race or sex) should be more straightforward. Whether the algorithm was given access to such a factor can be seen directly in the training procedure.

The problem is that even if the algorithm was not given access to information about race or sex, we can nonetheless have disparate treatment if the algorithm ends up favoring whites and men because of customer discrimination (as in case (5)) and statistical reality (as in case (6)). The real question is how to establish that fact. The analysis we have developed thus far suggests an approach to addressing this question, via the framework outlined above for making the training data, objective function, and resulting screening algorithm available for examination and experimentation. We can use access to these objects to try determining whether the facts are as stated in the two cases – much more effectively so than in cases (1) and (2).

In cases (7) and (8), the disparate impact issues should be equally straightforward. The existence of disparate impact is clear; the data will prove it, and also show its magnitude. We can also attempt to specify the practice that gives rise to those impacts. Then the question is the standard one: Can the disparate impact be justified, given the relevant standard? That is the same question that would be asked if an algorithm were not involved.

The presence of the algorithm goes further – it makes it possible to quantify the tradeoffs that are relevant to determining whether there is "business necessity" (or some other justification for disparate impact). Algorithms by construction produce not just a single ranking of applicants. They can produce a set of rankings that vary based on one's tolerance for disparate impact. For each of these rankings, they additionally quantify their effect on the objective, such as sales. This allows us to answer exactly the question, "What is the magnitude of the disparate impact, and what would be the cost of eliminating or reducing it?" In case (8) we can say, "What would be the effect on overall job performance if we were to reduce the disparate impact for every level of reduction?" The issue of the business necessity behind disparate impact now becomes easier to litigate, more readily separating specious from genuine arguments.

## E. Clarifying nondiscrimination versus affirmative action

Private and public institutions are motivated by two goals: ensuring nondiscriminatory behavior, regardless of gender, race, ethnicity, sexual orientation, or disability status, and helping members of disadvantaged groups – sometimes producing affirmative action, understood as preferential treatment for members of such groups. The latter is of course far more controversial than the former. On one view, affirmative action compromises the goal of nondiscrimination and is fatally inconsistent with it;[139] on another view, the two are compatible.[140] Algorithms help clarify the relationship between them.

Algorithms have *clearly stated objectives*. There is no need to guess whether an algorithmic rule is rank-ordering job applicants based on expected college GPA instead of by something else. Algorithms also let us precisely *quantify tradeoffs* among society's different goals. Suppose the university admissions committee cares about racial diversity and also about selecting a student body that would do as well as possible academically, as measured by grade point average (GPA) at (say) the end of the first year of college. An algorithm might give us the predicted GPA for every applicant, which would let us rank-order all applicants and count down the list until the target number of admissions is reached. But it would also be possible to create separate rank-ordered lists for African-American and white applicants, which would allow the university to count down the list of African-American applicants until whatever diversity target is hit. We can then calculate what happens to average GPA of the admitted class under that hypothetical admissions target. Put differently, with the algorithm it is now possible to see whether there is a tradeoff between diversity and academic performance of the incoming class, and if so, trace out what that tradeoff curve looks like.

There is no obvious answer to the question of whether the drop in average GPA is worth whatever benefits arise from improving racial diversity, and the algorithm should not be imagined to provide one. The key point is that the algorithm, by forcing us to specify what outcome we care about and providing us with a rank-ordered list of applicants by the predicted outcome, lets us precisely quantify what we would be giving up (if anything) to achieve these other objectives. We need a suitable legal framework to ensure that we can capitalize on this opportunity.

## F. Combating Discrimination

Regulation of the algorithm requires us to be able to identify and interrogate human choices in the construction of the training algorithm, and specifically their decisions about:

- What outcome to predict
- What inputs (candidate predictors) to make available to the algorithm for consideration
- The training procedure itself

As a result, an important requirement for a new legal framework to prevent discrimination through the use of algorithms is transparency. These specific questions that come out of our framework help make clear that for the successful regulation of algorithms, transparency is not an end in its own right but rather a means to an end. One might say that transparency is key for any legal system, including laws built for purely human decision-making: we need people to be transparent about what they are doing. But given the limits of human cognition, even a well-intentioned person will have difficulty "handing over" what is asked for.

[139] An influential discussion is John Hart Ely, *The Constitutionality of Reverse Racial Discrimination*, 41 U. CHI. L. REV. 723 (1974).
[140] *See* Charles Lawrence, *Two Views of the River: A Challenge of the Liberal Defense of Affirmative Action*, 101 COLUM. L. REV. 928 (2001).

In contrast, the inclusion of an algorithm in the decision loop now provides the opportunity for more feasible and productive transparency. Effective transparency is easier to imagine once an algorithm is involved, because each of the things we are now asking the firm to hand over – the choice of outcome, candidate predictors, training sample – is a tangible object. Each of them is, or at least could be, stored as part of the organization's standard daily operations.

The limits of this sort of transparency for many applications today is that several of the objects we would require for proper algorithmic regulation might not currently be stored by the relevant firm, particularly in online learning settings where data flow in at a massive volume (such as Web search or online ad delivery). Here is where the distinction becomes important between an engineering exercise of the sort data scientists normally consider as opposed to the sort of legal exercise in which we are engaged here.

In the design of a legal framework to deal with potential discrimination when algorithms are involved, there are many options. New statutes or regulations could change data and storage requirements imposed on algorithmic tasks like Web search to make them subject to the sort of discrimination inquiries we outline here, which in turn would expand the set of applications to which this framework applies. In the context of a specific legal challenge to alleged discrimination, it is not impossible that courts might insist on such requirements in order to test whether a violation has occurred (though that would be an admittedly unusual and aggressive step). The costs of compliance in this case could become quite large. Whether those costs are worth incurring will hinge on some assessment of whether the benefits justify them. The history of regulation includes many cases of large record-keeping costs being imposed on private firms in cases where the benefits were believed to be considerable; for example, after the financial market crash of 2008 almost brought about the collapse of the global economy, the Dodd-Frank act required firms to keep records on every single transaction that occurred in the $400 *trillion* swaps market.[141] That is a lot of transactions. In the abstract, it is not unreasonable to expect that if algorithms are being used, data and storage requirements could become automated at reasonable cost.

For regulating possible discrimination within the public sector in particular there is a different challenge – the risk of *broken procurement processes*.[142] Some of the organizations using algorithms to help inform critical decisions, particularly government agencies, lack the internal capacity to build and maintain this type of data-driven system.[143] So they understandably turn for help to some outside organization with the relevant expertise.[144] Those outside organizations are frequently for-profit firms. What they are selling is partly the implementation of their algorithm, but more importantly it is the algorithm itself. They – understandably, though not laudably – do not want to make the detailed workings of their algorithm public for fear of reducing the value of their intellectual property.

[141] *See, e.g.,* Dodd-Frank Act, Pub. L. No. 111-203, § 729, 83 Stat. 1376-2223 (2010) (prior to 1975 amendment).
[142] *See* Christopher Yukins, *The U.S. Federal Procurement System: An Introduction* (2017), available at https://papers.ssrn.com/sol3/papers.cfm?abstract_id=3063559.
[143] For relevant discussion, *see* Sarrah Chraibi et al., *An Exploratory Model for Outsourcing-purchasing Activities Based on a Comparative Study*, 3 SUPPLY CHAIN FORUM 138 (2017).
[144] *See id.*

For the same reason: Once some organization has incurred all the up-front costs of building and installing a new algorithmic system, the outside firm has leverage that tempts it to charge high prices for updating the tool. If the algorithm is not regularly updated there is the risk of what Koepke and Robinson call "zombie predictions," where the algorithm is built on an old dataset that reflects conditions that are quite different from when the predictions are now being applied.[145] Yet cash-strapped agencies too frequently cannot or will not pay their consultants what is needed to update the models.

## VI. USING ALGORITHMS TO PROMOTE EQUITY

Algorithms have the potential to help us to excise disparate treatment, to reduce discrimination relative to human decision-making, to limit disparate impacts, and also to predict much more accurately than humans can in ways that disproportionately benefit disadvantaged groups – what we call the "disparate benefit" of algorithms.

### A. De-bias relative to humans

As we have seen, the tendency to classify others into "in-groups" and "out-groups" is a key feature of human psychology that contributes to explicit and implicit biases throughout society. For example, Bertrand and Mullainathan show that resumes with a typically African-American name are much less likely to receive a call-back than similar resumes that list a common white name.[146] To overcome the effects of these human biases, we could try to exhort hiring managers to ignore irrelevant information on resumes. Or we could provide them with some sort of implicit bias training. But given the opacity of human decision-making, it would be very difficult to determine whether such efforts had actually been successful.[147]

The use of an algorithm is an alternative way to try to deal with the bias of human decision-making. To the algorithm, the name on a resume, race, age, sex, or any other applicant characteristic are candidate predictors like any other: variable X42. If this variable is not predictive of the outcome, the algorithm will not use it. And since predicting the outcome is *all* the algorithm is designed to do, we do not have to worry about any hidden agenda on the part of the algorithm itself. And recall that if use of a characteristic *is* predictive but would violate antidiscrimination law, use of that characteristic can and should be prohibited. But here there is an important qualification, to which we now turn.

### B. Access to the protected variable promotes equity for the algorithm

Consider a firm that is trying to decide which sales people to steer towards its most lucrative clients based on a prediction of their future sales level. Candidate predictors include (1) past sales levels and (2) manager ratings. Suppose that for men, managers provide meaningful assessments that include useful signals about employee performance that are not fully captured in the past sales data. But suppose that for women, the managers discriminate and give the lowest possible ratings. (This is of course disparate treatment and therefore unlawful.) An algorithm that is *prohibited* from knowing gender might well use manager ratings as a predictor, because it has a useful signal for half the sample. And because the algorithm in this case does not know who is male and who is female, it has no choice but to assume that manager ratings mean the same thing for all workers. The resulting predictions would understate the future productivity of women and hence contribute to gender gaps in earnings.

---

[145] *See* John Logan Koepke & David G. Robinson, *Danger Ahead: Risk Assessment and The Future of Bail Refor*m, 93 WASH. L. REV. 1725-1807 (2017).
[146] *See* Bertrand & Mullainathan, *supra* note.
[147] *See* S.M. Jackson et al., *Using Implicit Bias Training to Improve Attitudes Toward Women in STEM*, 17 SOCIAL PSYCHOLOGY OF EDUCATION 419 (2014).

But what happens if we instead allowed the algorithm to be *aware* of gender? With adequate training data, the algorithm could detect that manager ratings are predictive of future sales for men but not for women. Since the algorithm is tasked with one and only one job – predict the outcome as accurately as possible – and in this case has access to gender, it would on its own choose to use manager ratings to predict outcomes for men but not for women. The consequence would be to mitigate the gender bias in the data.[148] Clearly such a mitigating effect will not result from the use of a protected attribute in every situation, but we can easily find additional scenarios where similar considerations arise.

For example, allowing the algorithm to have access to protected-class membership can also promote what many people would consider to be equity in cases where the relationship between the candidate predictors and outcomes differ between the advantaged and disadvantaged groups. To use an admittedly more controversial example, suppose that we have two college applicants who both score 1,100 on the SAT. One of them is from New Trier, an affluent north-shore suburb of Chicago where the median family income is $145,000 (three times the national average) with public schools among the nation's best. The other is from Englewood, a south side Chicago neighborhood with median family income under $20,000 and among the city's highest homicide rates. Outside of extraordinarily unusual circumstances, it *cannot* be the case that the amount of effort, persistence, and extra learning on one's own required to score 1,100 on the SAT is the same for the student who starts with every possible advantage as for the one forced to overcome a long list of difficult obstacles.

Yet when we prohibit an algorithm from having access to information about the college applicant's disadvantaged group membership (in this case, let us stipulate, race as well as economic circumstances and background[149]), that is *exactly* what we are forcing the algorithm to assume. From the standpoint of current law, it is not clear that the algorithm can permissibly consider race, even if it ought to be authorized to do so; the Supreme Court allows consideration of race only to promote diversity in education.[150] Whatever the law requires, many people would find it tempting to say that while the algorithm should focus on factors independent of race (poor neighborhood, violent crime, and so forth), it should not focus on race itself.[151] But sometimes race itself is relevant as a predictor, and on plausible assumptions, there is nothing invidious about insisting on that point.

[148] As another example, consider a city in which half of all residents are white, half are black. Suppose that white residents and black residents engage in criminal activity at exactly the same rates, and that the police force in this city manages to arrest everyone who commits a crime. But in addition, the police department includes some discriminatory officers and as a result half of all arrests made to black residents are actually false arrests (that is, an arrest when the person was not reasonably thought to have committed a crime). In this case, the arrest rate will be higher for blacks than for whites even though true rates of crime are the same. Now suppose we use these data to predict failure to appear in court, or FTA, which (for the moment) we will assume is accurately measured. An algorithm blinded to race has no choice but to assume that the relationship between prior arrests and FTA risk is the same for both groups; since the rate of prior arrests is higher for blacks as whites, the race-blind algorithm would generally predict a higher FTA risk for blacks. In contrast, an algorithm able to use race would detect in the data that (in our stylized example) the effect of each prior arrest on FTA risk was smaller for black residents than for whites; with more prior arrests for blacks than whites, the net result would be equalized predicted FTA risks for blacks and whites by the race-aware algorithm.

[149] We acknowledge that the stipulation is controversial. To skeptics, we emphasize that it is a stipulation.

[150] *See* Grutter v. Bollinger, 539 U.S. 306 (2003).

[151] *See* Gratz v. Bollinger, 539 U.S. 244 (2003).

These are not just hypothetical examples; we can see them play out in actual data. In Kleinberg et al.,[152] we use data on a nationally representative sample of teens to predict college performance and simulate different admission decisions.[153] Table 1 shows that, consistent with past studies, high school outcomes differ on average by race, and as a result we see differences in average college outcomes as well. Figure 1 shows that a race-aware algorithm allows us *substantially to increase the share of admitted students who are minority* (holding average GPA constant) relative to ignoring information about race. The share of the admitted class that has GPA<2.75 is on the y-axis and the share that is African-American is on the x-axis. We show results from a race-aware algorithm, a race-blind algorithm, and an algorithm that pre-processes the data so that the average of all high school predictors is the same on average for white and black students.[154] The algorithms let us rank-order applicants by predicted GPA outcome.

If we do this separately by race, *we can select whatever share of the incoming class we wish to be minority by deciding how far down the minority student list to go*. The tradeoff curve is upward-sloping for each algorithm (more diversity, higher share GPA < 2.75), but the race-aware algorithm dominates the others. For example, if we wanted to hold the share of admitted students with GPA<2.75 at 14%, the share of admitted students who are black would be 7% with the race-blind model, 11% with the pre-processed model, and *19% with the race-aware algorithm*.

Figure 2 shows why the race-aware algorithm dominates. This "heat map" shows the share of black students in different predicted GPA "bins" according to the race-blind (x-axis) or race-aware (y-axis) algorithms. Observations that lie off of the 45-degree line show the models disagree. For example, the race-blind algorithm says the students in the right-hand lower corner have high likelihood of GPA<2.75 (9th decile), but the race-aware predictor (accurately) tells us they are actually in the *lowest-risk* decile. The race-blind model mis-ranks these students because the relationships between high school predictors of college success (such as test scores) turn out to be different for white versus black students.[155]

This discussion highlights how the mechanical application of current antidiscrimination law to algorithms might actually have harmful effects on exactly those populations we seek to protect. Our goal here is not to offer a final view on whether and when current law would forbid decision-makers – whether human beings or algorithms – from explicitly considering race. It is enough to say that courts would be very uncomfortable with that practice, and their discomfort, in the context of algorithms at least, might well be a mistake.

### C. Disparate benefit from improved prediction

---

[152] The dataset we use is the US Department of Education's National Education Longitudinal Study of 1988, which captures information on a nationally representative sample of students who entered 8th grade in the fall of 1988. U.S. DEPARTMENT OF EDUCATION, NATIONAL EDUCATION LONGITUDINAL STUDY OF 1988 (1988). Follow up surveys were conducted in 1990, 1992, 1994 and in 2000 (when respondents were in their mid-20s). *See Available Data*, U.S. DEPARTMENT OF EDUCATION, https://nces.ed.gov/surveys/nels88/data_products.asp. We limit our analysis sample to those who participated in the 2000 survey wave, and had ever attended a four-year post-secondary institution. To simplify the analysis, we focus on comparing just non-Hispanic white students (N=4,274) with black students (N=469). In the public-use version of the NELS we use here, we do not have access to ACT or SAT scores. But we do have the results of how students did on standardized academic achievement tests that the NELS administered to students in four academic areas: math, reading, science, and social studies.

[153] *See* Kleinberg et al., *supra* note.

[154] We might do this if, for example, we believed that systematic differences across groups in the average values of the predictor variables (Xs) was due to some sort of societal unfairness, or to biased data.

[155] When we use the race-blind algorithm, we force the statistical model to assume the relationship is the same for both whites and blacks, which distorts the predicted outcomes for black applicants. This leads us inadvertently to say some high-performing black applicants are actually low-performing, that is, to understate their academic potential and hence reject them, while simultaneously admitting lower-performing black applicants instead.

There is a less controversial but comparably important point. The largest potential equity gains may come from simply predicting more accurately than humans can. This increase in accuracy can generate benefits that disproportionately accrue to disadvantaged groups, leading to what we call "disparate benefit."

For example, landlords in the private housing market have difficulty predicting which tenants are at high risk for skipping rent payments. This difficulty may lead landlords to implement blanket rules, such as requiring first and last month rent to protect against missed payments. For affluent renters, this is a minor inconvenience. But for low-income, low-wealth families, this can be the difference between leasing an apartment and doubling up with someone else (or even becoming homeless). Better prediction could allow landlords to relax collateral requirements, which would *change* the system in ways that helps disadvantaged families.

Better prediction could also let us *expand* some systems in ways that help disadvantaged groups. Compared to affluent white families, low-income minority families tend to live in neighborhoods that do not only have higher poverty rates, but are also located further from good schools, health care and jobs.[156] This distance is particularly a problem for families who rely on public transportation, which in the US remains, as Pendall puts it, "slow, inconvenient, and [lacks] sufficient metropolitan-wide coverage to rival the automobile."[157] While car loans have increased in recent years, interest rates often remain high in part because of non-payment: 6 million people are at least 90 days behind on their car loans.[158] High interest rates surely contribute to differences in car ownership by income and employment status.[159] Better prediction of payment risk could allow lenders to lower interest rates for low-risk families and expand access to car ownership in ways that disproportionately benefit the disadvantaged.

We are offering somewhat conjectural examples, but these potential gains are not merely conjectural. Consider an application where better prediction lets us shrink a system that disproportionately harms disadvantaged groups: pre-trial release decisions for criminal defendants.[160] State law in New York requires judges make these decisions based on a prediction of defendant risk of failure to appear (FTA) in court in the future. We use data from New York City on all cases that were continued at arraignment over a five-year period and build a model to predict FTA risk. The predictor variables in the model consist of age (a legally allowable variable), current offense, and prior criminal record. The data show that the judges' risk predictions (implicit in their release decisions) are correlated with the algorithm, but that compared to the algorithm, the judges make numerous serious mistakes: they detain many low-risk people and release many high-risk ones.[161]

---

[156] Rolf Pendall, *For Many Low-Income Families, Car May Be Key to Greater Opportunity,* URBAN INSTITUTE (2014), https://www.urban.org/urban-wire/many-low-income-families-cars-may-be-key-greater-opportunity.

[157] *Ibid.*

[158] *See* Gwynn Guilford, *American Car Buyers Are Borrowing Like Never Before – And Missing Plenty Off Payments, Too,* QUARTZ (Feb. 21, 2017), https://qz.com/913093/car-loans-in-the-us-have-hit-record-levels-and-delinquencies-are-rising-fast-too/.

[159] *See* Steven Raphael & Lorien Rice, *Car Ownership, Employment, and Earnings,* 52 J. OF URBAN ECON. 109 (2002).

[160] *See* Kleinberg, *supra* note.

[161] In particular, judges seem substantially to over-weight the severity of the current charge (i.e., whatever offense the person was arrested for that led them to wind up in court). *See id.*; Cass R. Sunstein, *Algorithms, Correcting Bias*, SOCIAL RESEARCH (forthcoming 2019). When the judges form their own implicit list of defendants to prioritize for detention based on their subjective assessment of defendant risk, the judges wind up putting too many felony arrestees towards the top of their list and too many misdemeanor arrestees towards the bottom of the list. But the machine learning algorithm makes clear that prior record matters a lot in terms of future risk, and human judges are not fully taking this into account. Kleinberg et al., *supra* note.

If we made these decisions using the algorithm's risk prediction rather than those of the judge, we could prioritize detention just for those people who are high risk. Indeed, some of the high-risk people judges release are *so* high risk that if we detained them, we could release multiple low-risk people without increasing FTAs. Indeed, if we followed the algorithm's release recommendations, we could reduce the jail population by 42% without increasing FTA rates at all.[162]

Figure 3 provides a more concrete sense for what this means in practice. In a city in which around 50,000 people spend time in jail each year, it would mean about 20,000 fewer people spending time behind bars on Riker's Island each year.[163] This would be the equivalent of closing Riker's Island at the end of July every year, with no increase in FTA or re-arrests. And note who benefits the most from this large reduction in detentions: the two groups who together account for nearly 90% of all current jail spells -- African-American and Hispanic defendants.[164] This is a disparate benefit, and an especially large one.

### D. Algorithms can reveal our own biases

A final potential benefit arises from the fact that algorithms can help reveal our own biases. Imagine a large, growing firm that is inundated with job applications. To help prioritize which resumes to consider as it hires and expands, the firm builds an algorithm to rank applications. The outcome predicted by this algorithm is based on the people the firm's managers hired in the past. The firm then notices that the new algorithm mostly hires men. A possible response would stem from the recognition that the algorithm, given how it was constructed, reveals implicit or explicit biases among the firm's HR team. In that light, the firm might try to overcome biased human judgments by building a new algorithm that predicts an outcome less infected by human bias, such as a more objective measure of worker productivity once hired. A perhaps less helpful response would lose sight of what is responsible for the algorithm's bias, and drop the algorithm altogether to revert back to a purely human hiring process.[165]

The transparency of algorithms will have other consequences that might be uncomfortable for many people. Recall that the disparate impact standard forbids disproportionate adverse effects on members of certain groups, unless there is a strong independent justification for the requirement or practice that creates those adverse effects. Here we can see the lines blurring between antidiscrimination principles and affirmative action.[166] Suppose that we have two candidate algorithms to predict worker productivity. One of them would lead to hiring a set of workers that is 1% more productive than those hired by the other, but reduces the number of minorities hired by 10%. Is this productivity gain large enough to provide "strong independent justification" for that algorithm? Answering that question unavoidably winds up requiring value judgments about how to trade off avoidance of disparate racial impacts against other social objectives (such as output).

[162] *Id.*

[163] *See* NEW YORK CITY DEPARTMENT OF CORRECTION, NYC DEPARTMENT OF CORRECTION AT A GLANCE (2018), https://www1.nyc.gov/assets/doc/downloads/press-release/DOC_At%20a%20Glance-entire_FY%202018_073118.pdf.

[164] These potential gains are not limited to pre-trial release decisions. *See* Aaron Chalfin et al., *Productivity and Selection of Human Capital with Machine Learning*, 106 AM. ECON. REV.: PAPERS & PROCEEDINGS 124 (2016). Using algorithmic predictions rather than human judgment can help police departments hire officers at lower risk for adverse outcomes like police-involved shootings, complaints for verbal abuse, or complaints for physical abuse, and can help school systems hire more effective teachers. Because police misconduct so disproportionately involves disadvantaged populations, *see* Roland Fryer, *Reconciling Results on Racial Differences in Police Shootings,* 108 AM. ECON. ASSOC. PAPERS & PROCEEDINGS 228 (2018), and similarly because the most effective teachers tend to prefer working with the most affluent students leaving the least effective to serve disproportionately low-income and often predominantly minority schools, *see, e.g.*, Mimi Engel et al., *New Evidence on Teacher Labor Supply*, 51 AM. EDUC. RES. J. 36 (2014), any system for improving the average quality of police officers and teachers will have particularly large benefits for low-income and minority populations.

[165] Interestingly, this is what some media accounts suggest may have happened in 2018 at Amazon. *See* Reuters, *Amazon Ditched AI Recruiting Tool That Favored Men for Technical Jobs,* THE GUARDIAN, (Oct. 11, 2018), https://www.theguardian.com/technology/2018/oct/10/amazon-hiring-ai-gender-bias-recruiting-engine.

[166] See Strauss, *supra* note.

Another way to put it is to say that those who favor affirmative action programs are sometimes willing to sacrifice some value for the sake of other goals (such as racial justice as they see it). Those who want to prevent disparate impacts, or who want to ensure that any disparate impact is strongly justified, are willing to do *exactly the same thing*. In practice, current antidiscrimination law says that it is worth suffering something, but not too much, to stop disparate impacts on (say) African-Americans or women. But how much? Algorithms permit unprecedented clarity about these questions by allowing us to specify the magnitude of tradeoffs. What if, instead of a 1% productivity gain for a 10% decline in minority hiring, it was a 10% productivity gain for a 1% decline in minority hiring? Many people might respond that the relevant judgments will depend on the precise numbers, which are typically impossible to quantify with "black box" human decision-making.

## VII. Conclusion

It is tempting to think that human decision-making is transparent and that algorithms are opaque. We have argued here that with respect to discrimination, the opposite is true. The use of algorithms offers far greater clarity and transparency about the ingredients and motivations of decisions, and hence far greater opportunity to ferret out discrimination.

For those who wish to reduce discriminatory behavior, this is a massive opportunity. Countless decisions have the flavor of the screening problems we discuss here; they hinge on a prediction. There is powerful evidence of discrimination in current human decision-making. To offer just a few examples:

- Audit studies that randomly assign otherwise-equivalent white and black applicants to apply at different firms find that white applicants are called back at more than twice the rate of black applicants, 34% versus 14%.[167] Reducing this bias would do an enormous amount of social good given there are over 6 million job openings in the US at any point in time.[168]
- Audit studies of the US housing market, which originates $2 trillion in new mortgages each year,[169] find that minority borrowers are treated differently from and worse than white borrowers.[170]
- Bias also arises in the health sector, which accounts for $3.5 trillion in spending each year in the US alone (equal to 18% of GDP).[171] For example, when doctors were shown two equivalent

[167] This figure is for whites and blacks without criminal records. Among those with criminal records we see similarly large differences in call-back rates, equal to 17% versus 5%, respectively. *See* Devah Pager, *The Mark of a Criminal Record*, 108 AM. J. SOC. 937 (2003).
[168] Bureau of Labor Statistics, *Economic News Release: Job Openings and Labor Turnover Summary*, (Jan. 8, 2019), https://www.bls.gov/news.release/jolts.nr0.htm.
[169] *See* Tendayi Kapfidze, *U.S. Mortgage Market Statistics: 2018,* MAGNIFY MONEY BY LENDING TREE (Dec. 21, 2018), https://www.magnifymoney.com/blog/mortgage/u-s-mortgage-market-statistics-2017/.
[170] *See* Margery Austin Turner et al., *All Other Things Being Equal: A Paired Testing Study of Mortgage Lending Institutions – Final Report* (2002).
[171] CENTERS FOR MEDICARE & MEDICAID SERVICES, NATIONAL HEALTH EXPENDITURE DATA FOR 2017 (2018), https://www.cms.gov/research-statistics-data-and-systems/statistics-trends-and-reports/nationalhealthexpenddata/nhe-fact-sheet.html; *GDP*, THE WORLD BANK, https://data.worldbank.org/indicator/NY.GDP.MKTP.CD.

patient histories, the chances of recommending a beneficial procedure (cardiac catheterization) were 40% lower for women and minorities than white males.[172]

For each of these critically important decisions, more and more data are becoming available over time on how people in the past made these decisions, the characteristics of these people and their decision-making environments, and the consequences of different decisions. These data make it increasingly possible to build data-driven statistical prediction models. Those models might be used to detect, reduce, or eliminate existing discrimination.

Algorithms have extraordinary promise. They have the *potential* to make important strides in combating discrimination, at least as the legal system has long understood it. But principles of transparency and auditability, fair and nondiscriminatory choice of data, and reasonable algorithmic objective are essential, not least to help understand and select the tradeoffs that people use algorithms to make.

[172] *See* Kevin Schulman et al., *The Effect of Race and Sex on Physicians' Recommendations for Cardiac Catheterization*, 340 NEW ENGLAND J. OF MED. 618 (1999).

## I. APPENDIX: A CHARACTERIZATION OF ALGORITHMIC BIAS

In this appendix, we justify the claim that algorithmic bias can be decomposed completely into three components: bias in the choice of input variables, bias in the choice of outcome measure, and bias in the construction of the training procedure. The disparity that remains after accounting for these three forms of bias corresponds to the structural disadvantage of one group relative to another.

To make this claim precise, we use the following formalism. Suppose that in a screening problem we have applicants from two groups: an advantaged group and a disadvantaged group. The complete description of each applicant is specified by a (long) feature vector $x$, and their true productivity is specified by a function $f(x)$ of this feature vector.

Suppose that in the advantaged group, a $p(x)$ fraction of the applicants has feature vector $x$. In other words, a random applicant drawn from the advantaged group will have feature vector $x$ with probability $p(x)$. Analogously, in the disadvantaged group, a $q(x)$ fraction of the applicants have feature vector $x$. Thus, the average productivity of applicants in the advantaged group is the sum over all $x$ of $p(x)f(x)$, which we will write more compactly as $\sum_x p(x)f(x)$, and the average productivity of applicants in the disadvantaged group is the corresponding sum $\sum_x q(x)f(x)$.

We add one more piece of notation: for any function $v$, we let $D(v)$ denote the difference in the average value of $v$ between the advantaged and disadvantaged groups; that is, $D(v) = \sum_x [p(x) - q(x)]v(x)$. Thus, the difference in average productivity between the two groups is equal to $D(f)$; we can think of this as quantifying the structural disadvantage of one group relative to the other.

Now, the designer of an algorithm for the screening problem does not have access to the true function $f$, nor to an applicant's full feature vector $x$. Instead, the algorithm designer creates a screening rule following the process described in the paper.

- The designer specifies the performance of each applicant using a function $g(x)$, which will in general be different from $f(x)$.
- For each applicant, the designer does not use the full feature vector $x$, but instead a reduced representation $r(x)$. This means that neither the function $f$ nor $g$ can be directly applied to the representation of an applicant; rather, a function $h$ is applied to $r(x)$, yielding a value $h(r(x))$. We write $h \circ r$ for the composition of the functions $h$ and $r$ (obtained by first applying $r$, and then applying $h$); in this notation, the value computed for an applicant with true feature vector $x$ is $(h \circ r)(x)$.
- Finally, the designer must estimate $h$ from the training data they have collected, and this results in a function $t$. Like the function $h$, this function $t$ can be applied to the representation $r(x)$, yielding a value $(t \circ r)(x)$.

The result of this design process is an algorithm that, to an applicant with feature vector $x$, assigns a score of $(t \circ r)(x)$. The applicants are then ranked by this score, with the highest-scoring applicants selected.

The difference in average scores assigned to the advantaged and disadvantaged groups is thus $D(t \circ r)$, whereas the actual level of structural disadvantage is $D(f)$. How should we understand the contrast between these two quantities? The key idea is to write $D(t \circ r)$ in the following extended form:

$$D(t \circ r) = D(f) + (D(g) - D(f)) + (D(h \circ r) - D(g)) + (D(t \circ r) - D(h \circ r))$$

It is easy to verify that the left- and right-hand sides are equal. The advantage of the extended formulation on the right-hand side is that each of its form terms is directly interpretable, as follows.

- The first term, $D(f)$, is the average difference in $f$ between the two groups, and hence corresponds to the structural disadvantage.
- The second term, $D(g) - D(f)$, is the additional bias introduced by the choice of outcome measure: $g$ instead of $f$.
- The third term, $(D(h \circ r) - D(g))$, is the additional bias introduced by the choice of input variables: the reduced representation $r(x)$ instead of the full feature vector $x$.
- The fourth term, $D(t \circ r) - D(h \circ r)$, is the additional bias introduced by the training procedure: the fact that we must use an estimated $t$ rather than $h$.

This equation thus formalizes the sense in which, after accounting for three forms of bias introduced in the design of the algorithm – the choice of outcome measure, the choice of input variables, and the design of the training procedure – what remains is the level of structural disadvantage between groups. As discussed earlier, we may wish to pursue interventions aimed at reducing all three forms of bias as well as the structural disadvantage, but it is important to understand how our interventions are distributed across these different effects.

Table 1
Descriptive statistics from NELS dataset for predicting college performance

| | Whites | Blacks |
|---|---|---|
| N | 4,274 | 469 |
| College outcomes (2000 follow-up) | | |
| Graduate with BA by 2000 | 67.4% | 50.9% |
| College GPA ≥ 2.75 | 82.2% | 69.5% |
| College GPA ≥ 3.25 | 48.4% | 31.1% |
| Female | 53.2% | 56.7% |
| 10th grade NELS test scores (standard deviation = 10) | | |
| Reading | 56.3 | 50.0 |
| Math | 57.1 | 49.5 |
| Science | 56.2 | 47.5 |
| History | 56.0 | 50.0 |
| 10th grade course grades | | |
| English: Mostly A's | 29.4% | 19.6% |
| Math: Mostly A's | 26.9% | 14.9% |
| Science: Mostly A's | 26.9% | 14.4% |
| History: Mostly A's | 29.1% | 15.7% |
| 10th grade extracurricular hours per week | | |
| None | 21.4% | 30.2% |
| 0-1 | 16.9% | 14.4% |
| 1-4 | 21.9% | 23.4% |
| 5-9 | 16.3% | 13.6% |
| 10-19 | 17.7% | 11.3% |
| 20+ | 2.1% | 1.2% |
| 12th grade NELS test scores (standard deviation = 10) | | |
| Reading | 56.0 | 50.0 |
| Math | 57.1 | 49.5 |
| Science | 56.2 | 47.5 |
| History | 56.0 | 50.0 |
| Number of credits taken by 12th grade by subject: | | |
| English | 3.6 | 3.4 |
| Math | 3.2 | 2.9 |
| Science | 2.9 | 2.5 |
| Social Studies | 3.1 | 2.8 |
| Took SAT | 60.4% | 63.1% |
| Took ACT | 52.6% | 37.9% |

Figure 1
Fairness vs. efficiency tradeoffs in college admissions using race-aware vs. race-blind algorithms

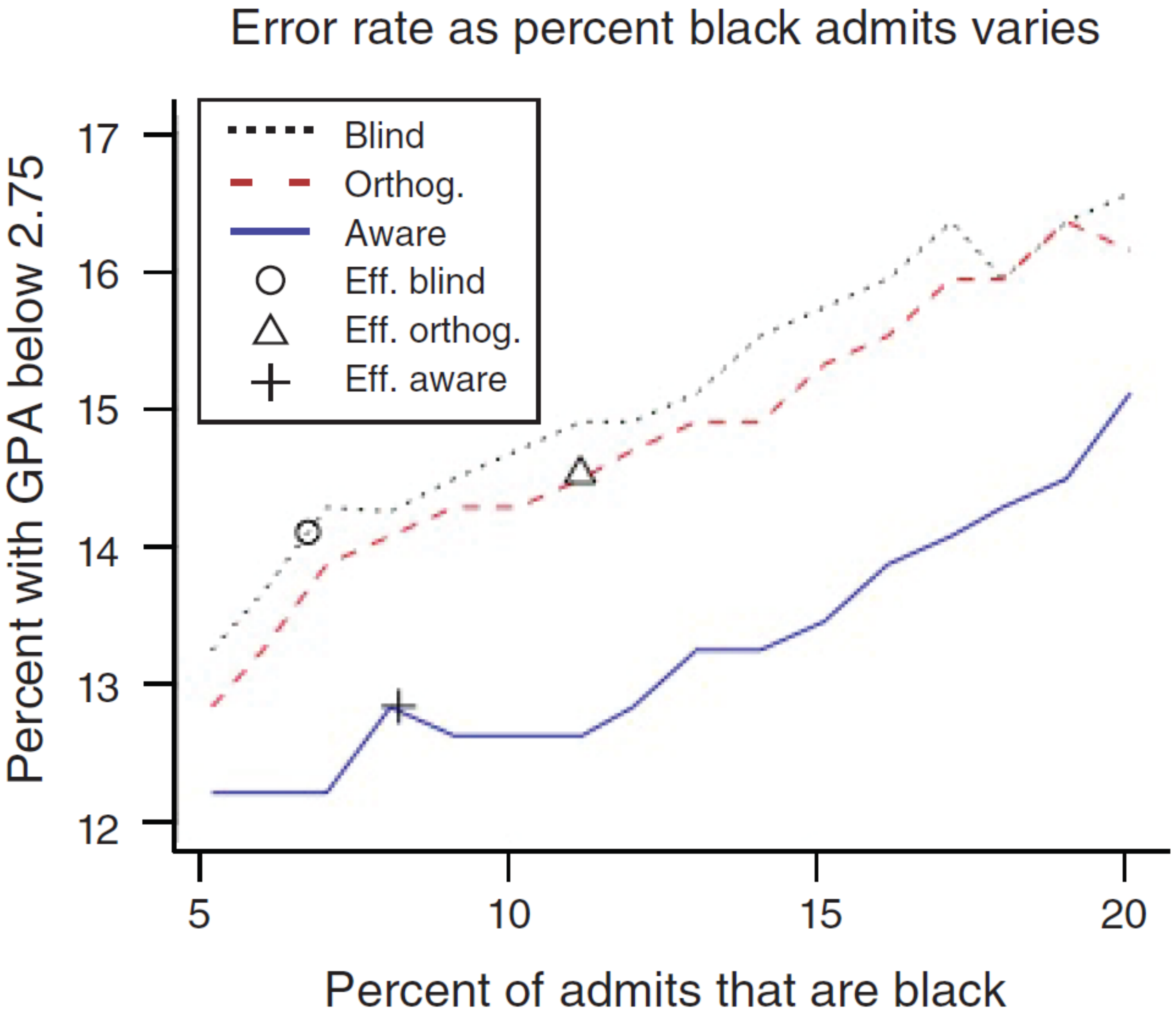

Source: Results from Kleinberg, Ludwig, Mullainathan and Rambachan (2018) using data from the NELS:88 to predict college performance for those students who attended a four-year college (as measured by GPA<2.75). The graph shows the results of rank-ordering all students by predicted performance and simulating an admission rule that admits the top half; the y-axis shows the percent of the "admitted class" that goes on to have a GPA below 2.75 ("mostly B's") while the x-axis shows the share of the admitted students who are African-American. The points in the graph show the result of rank-ordering applicants using linear probability models that (a) are "blinded" to any information about each applicant's race [circle]; (b) use information about each applicant's race to pre-process the data and orthogonalize the predictors to race, so that the average value of high school grades, test scores, etc. are now forced to be equal for black and white applicants [triangle]; and (c) use information about applicant race to form the prediction model [cross]. For each candidate algorithm, we also show what happens if we use our decision-making framework to achieve a target fairness level (share of the admitted college class that is African-American). Specifically, we take the predictions from a given algorithm, create a list of white applicants rank-ordered by predicted college performance, and a different list of black applicants rank-ordered by their predicted college performance, then work down the list of black applicants until we hit the target for minority enrollment in the hypothetical "admitted class." We can see that admission decisions using the race-aware algorithm dominate those from the race-blind or the race-orthogonalized algorithms, in the sense that for any given level of academic performance of the incoming freshman class, the race-aware algorithm lets us substantially increase the share of the admitted class that is African-American.

Figure 3
Mis-ranking of African-American applicants by race-blind algorithm

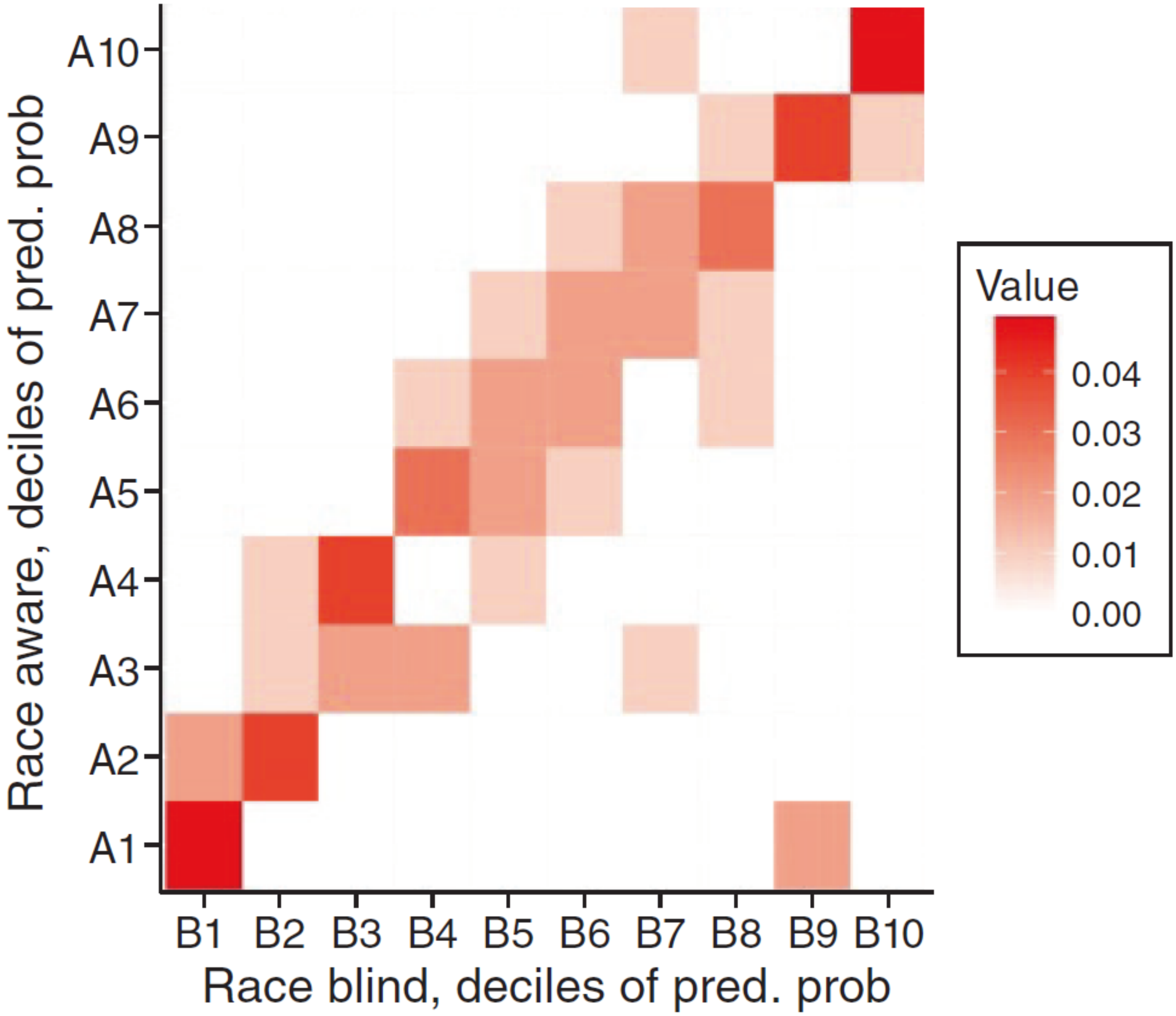

Source: Kleinberg, Ludwig, Mullainathan and Rambachan (2018). Using data from the NELS:88 dataset, we first predict for each observation their predicted college performance (measured as Y=1 if GPA<2.75) using an algorithm that is blinded to applicant race, and then again using an algorithm that has access to each applicant's race. We then take the sample of black students in the NELS:88 and use their predicted values to bin them into deciles based on the race-blind predictions (x-axis) and race-aware predictions (y-axis). If the two models rank-ordered everyone the same way, all the data would be along the 45-degree line. The "off diagonals" in the figure show mis-ranking.

Figure 3
Pre-trial detention rates in New York City under current human (judge) decisions versus algorithmic release rule that holds failure to appear (FTA) rate constant at current level

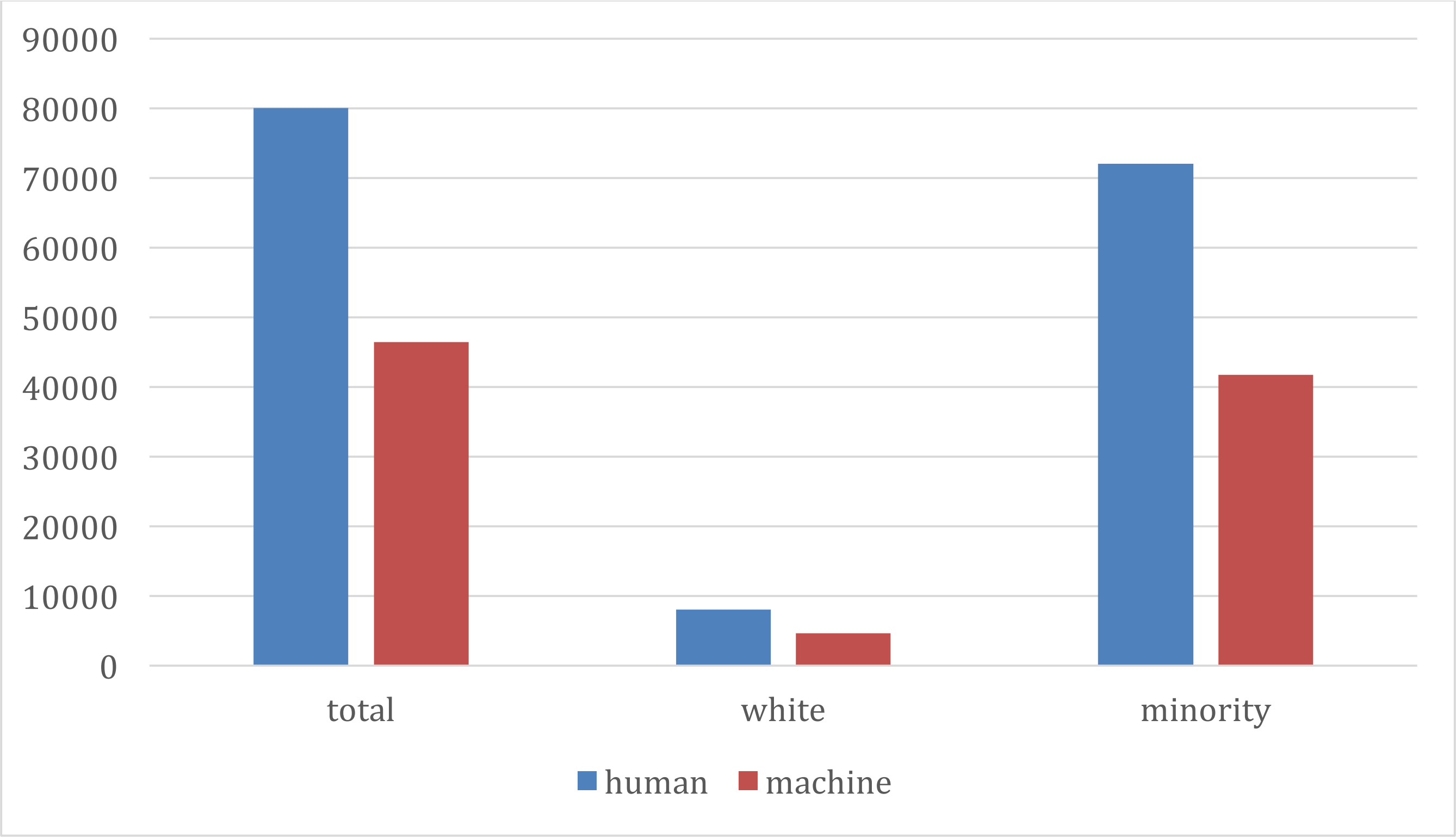

Source: Kleinberg, Lakkaraju, Leskovec, Ludwig and Mullainathan (2018). "Human decisions" represent current outcomes from the existing criminal justice system, which technically combines the input of judges, district attorneys, public defenders and a six-item risk check-list created by a local non-profit, the Criminal Justice Agency. "Machine decisions" represent the hypothetical detention outcomes if one were to make pre-trial detention versus release decisions using a machine learning algorithm (gradient boosted trees) to predict defendant risk, rank order defendants by risk, and detain only the number of defendants needed to keep the failure to appear rate constant at current levels.